\documentclass[12pt,preprint]{aastex}

\def\copernicus{{\it Copernicus\/}}
\def\imaps{{\it IMAPS\/}}
\def\fuse{{\it FUSE\/}}                 

\def\hst{{\it HST\/}}
\def\lya{Ly$\alpha$}
\def\lyb{Ly$\beta$}
\def\EE#1{\times 10^{\small#1}}
\def\chisq{$\chi^2$}

\def\teff{$T_{\rm{eff}}$}
\def\iue{{\it IUE\/}}
\def\euve{{\it EUVE\/}}

\def\mfarcs{\hbox{$~\!\!^{\prime\prime}$}}

\shorttitle{Deuterium in GD 246, WD 2331$-$475, Lan 23, and HZ 21.}
\shortauthors{Oliveira et al.}

\received{2002 October 31}
\begin{document}

\title{Interstellar Deuterium, Nitrogen, and Oxygen Abundances Toward GD 246, WD 2331$-$475, HZ 21, and Lan 23: Results from the {\bf \fuse}~Mission\altaffilmark{1}.}
%{Based on observations made with the NASA-CNES-CSA Far Ultraviolet Spectroscopic Explorer. {\it FUSE} is operated for NASA by the Johns Hopkins University under NASA contract NAS5-32985.}}

\author{Cristina M. Oliveira\altaffilmark{2}, Guillaume~H\'{e}brard\altaffilmark{3}, J.~Christopher~Howk\altaffilmark{2,4}, Jeffrey~W.~Kruk\altaffilmark{2}, Pierre~Chayer\altaffilmark{2,5}, and H.~Warren~Moos\altaffilmark{2}}
  
\altaffiltext{1}{Based on observations made with the NASA-CNES-CSA {\it Far Ultraviolet Spectroscopic Explorer}. \fuse~is operated for NASA by The Johns Hopkins University under NASA contract NAS5-32985.}
\altaffiltext{2}{Department of Physics and Astronomy, The Johns Hopkins University, Baltimore, MD 21218}
\altaffiltext{3}{Institut d\'{}Astrophysique de Paris, 98$^{bis}$ boulevard Arago, F-75014 Paris, France}
\altaffiltext{4}{Current Address: Center for Astrophysics and Space Sciences, University of California at San Diego, C-0424, La Jolla, CA, 92093}
\altaffiltext{5}{Primary affiliation: Department of Physics and Astronomy, University of Victoria, P.O. Box 3055, Victoria, BC V8W 3P6, Canada }

\begin{abstract} % OK OK
The interstellar abundances of D I, N I, and O I in the local ISM are studied using high-resolution spectra of four hot white dwarfs. The spectra of GD 246, WD 2331$-$475, HZ 21, and Lan 23 were obtained with the {\it Far Ultraviolet Spectroscopic Explorer} (\fuse) in the wavelength range 905~--~1187~\AA. The line of sight to GD 246 probes the Local Interstellar Cloud and at least one
 other H I cloud inside the Local Bubble, which contains most of the gas seen along this line of sight.
The column densities of H I, C II*, S II, and Si II are measured using archival {\it Hubble Space Telescope} STIS echelle-mode observations.
The H I column density is determined by fitting the strong damping wings of interstellar \lya~using a model atmosphere to account for the stellar continuum.
The sightline-averaged ratios for GD 246 are: D~I/H~I~=~(1.51~$\pm~^{0.39}_{0.33})\EE{-5}$, O~I/H~I~=~(3.63~$\pm~^{0.77}_{0.67})\EE{-4}$, and D~I/O~I~=~(4.17~$\pm~^{1.20}_{1.00})\EE{-2}$  (uncertainties are 2$\sigma$). This line of sight provides the fourth reliable \fuse~measurement of the Local Bubble D/H ratio. For the WD 2331$-$475 line of sight we find sightline-averaged ratios: D~I/O~I~=~(5.13~$\pm~^{2.20}_{1.69})\EE{-2}$ and D I/N I~=~(4.57~$\pm~^{1.88}_{1.45})\EE{-1}$. Toward HZ 21 the sightline-averaged ratios are: D~I/O~I~=~(4.57~$\pm~^{2.22}_{1.63})\EE{-2}$ and D I/N I~=~(4.27~$\pm~^{1.96}_{1.44})\EE{-1}$. In the higher column density sightline to Lan 23 the sightline-averaged ratios are: D~I/O~I~=~(3.24~$\pm~^{3.27}_{2.06})\EE{-2}$ and D~I/N~I~=~(3.16~$\pm~^{1.56}_{1.23})\EE{-1}$. Molecular hydrogen, corresponding to rotational levels J $\le$ 3, is clearly seen along this line of sight. No reliable H I measurements are available for WD 2331$-$475, HZ 21, or Lan 23. We combine the different abundance ratios computed here with previous published values to produce revised \fuse~abundance ratios for D I/H I, O I/H I, N I/H I, D I/N I, D I/O I, and O I/N I.

\end{abstract}



\section{INTRODUCTION} % OK OK

The present day abundance ratio of deuterium to hydrogen places important constraints on Big Bang nucleosynthesis (BBN) and the chemical evolution of galaxies. Since it is believed that deuterium is only produced in appreciable amounts in primordial BBN and destroyed in stellar interiors, the measurement of D I/H I in the interstellar medium (ISM) places a lower limit on the primordial abundance of deuterium. By comparing the ISM abundance of deuterium to its abundance in high-redshift intergalactic gas we should be able to understand better the effects of astration and chemical evolution of galaxies.
Measurements of the D/H ratio in intervening clouds of gas seen toward distant quasars have yielded a range of values D/H = (1.65 -- 4.0)$\EE{-5}$ \citep[and references therein]{2001ApJ...552..718O,2001ApJ...560...41P,2002ApJ...565..696L}.  
Measurements of D/H in the local ISM have been made with~\copernicus~\citep[e.g.][]{1973ApJ...186L..95R}, \hst, \citep[e.g.][]{1995ApJ...451..335L},~\imaps,~\citep{1999ApJ...520..182J,2000ApJ...545..277S}, and more recently~\fuse~\citep[and references therein]{2002ApJS..140....3M}.
A nearly constant ratio of D/H = $(1.5~\pm~0.1)\EE{-5}$ (1$\sigma$~on the mean) has been obtained in the Local Interstellar Cloud (LIC) by \citet{1998SSRv...84..285L}; recent measurements inside the Local Bubble \citep{2002ApJS..140....3M} appear to be consistent with a single value for D/H in the Local Bubble. Other measurements \citep{1979ApJ...229..923L,1983ApJ...264..172Y,1999A&A...350..643H,1999ApJ...520..182J,2000ApJ...545..277S} suggest variations of the interstellar D/H ratio beyond the Local Bubble, at the distance of a few hundred parsecs.

Until \fuse~was launched (1999 June 24) only the UV spectrographs onboard \hst~could be used for systematic studies of the deuterium abundance in the ISM. 
Since only the \lya~transitions of H I and D I are observable in the \hst~bandpass, such studies were restricted to low column density sightlines so that D I was neither completely obscured by the adjacent H I \lya~nor heavily saturated. With \fuse, we now have access to the complete Lyman series of deuterium (except \lya), which allows the study of a larger range of environments and provides tighter constraints on the abundance of D I compared to studies using only \lya. In addition, the numerous lines of N I, O I, and Fe II in the \fuse~bandpass can be used to trace the metallicity and dust content of the absorbing gas.

%A reliable H I column density cannot be derived from the \fuse~data, as all the available H I transitions typically lie on the flat part of the curve of growth. This is not true in cases where the column density of H I is so high that the \lyb~absorption line exhibits radiation damping wings. 
%Unfortunately these cases present several problems (such as a high molecular hydrogen content) that make this measurement very difficult.

In this work we present the first measurements of deuterium absorption toward the white dwarfs GD 246, WD 2331$-$475, HZ 21, and Lan 23. Table \ref{stellar_prop} summarizes the stellar parameters. This study adds four more sightlines to the seven sightline studies that constitute the first set of \fuse~deuterium measurements \citep[see overview paper by][and references therein]{2002ApJS..140....3M}. This paper is organized as follows. The observations and data processing are presented in \S 2; the analysis methodology is described in \S 3. The GD 246 line of sight analysis is presented in \S 4. The WD 2331$-$475, HZ 21, and Lan 23 lines of sight are analyzed in \S 5, and systematic effects are discussed in \S 6. We discuss the results in \S 7.

%{\large $\Longrightarrow$ INSERT TABLE 1}

\section{OBSERVATIONS AND DATA PROCESSING} % OK OK

\subsection{\fuse~Observations}

The \fuse~observatory consists of four co-aligned prime-focus telescopes and 
Rowland-circle spectrographs that produce spectra over the wavelength range 
905 -- 1187~\AA~with a spectral resolution of $\sim$~15~--~20~km~s$^{-1}$ (wavelength dependent), for point sources. Two of the optical channels employ SiC coatings, providing reflectivity in the
wavelength range $\sim$~905~--~1000~\AA,~while the other two have LiF coatings
for maximum sensitivity above 1000~\AA. Dispersed light is focused onto two
photon-counting microchannel plate detectors. With this arrangement of optical channels (LiF 1, LiF 2, SiC 1, and SiC 2) and detector segments (1A, 1B, 2A, 2B) the \fuse~instrument has 8 segments: LiF 1A, LiF 1B, LiF 2A, LiF 2B, SiC 1A, SiC 1B, SiC 2A, and SiC 2B. Four channels cover the wavelength range 1000~--~1080~\AA~while two channels each cover the ranges 900~--~1000~\AA~ and 1080~--~1180~\AA. The \fuse~mission, its planning, and on-orbit performance are discussed by \citet{2000ApJ...538L...1M} and \citet{2000ApJ...538L...7S}

\begin{figure}
\epsscale{0.80}
\plotone{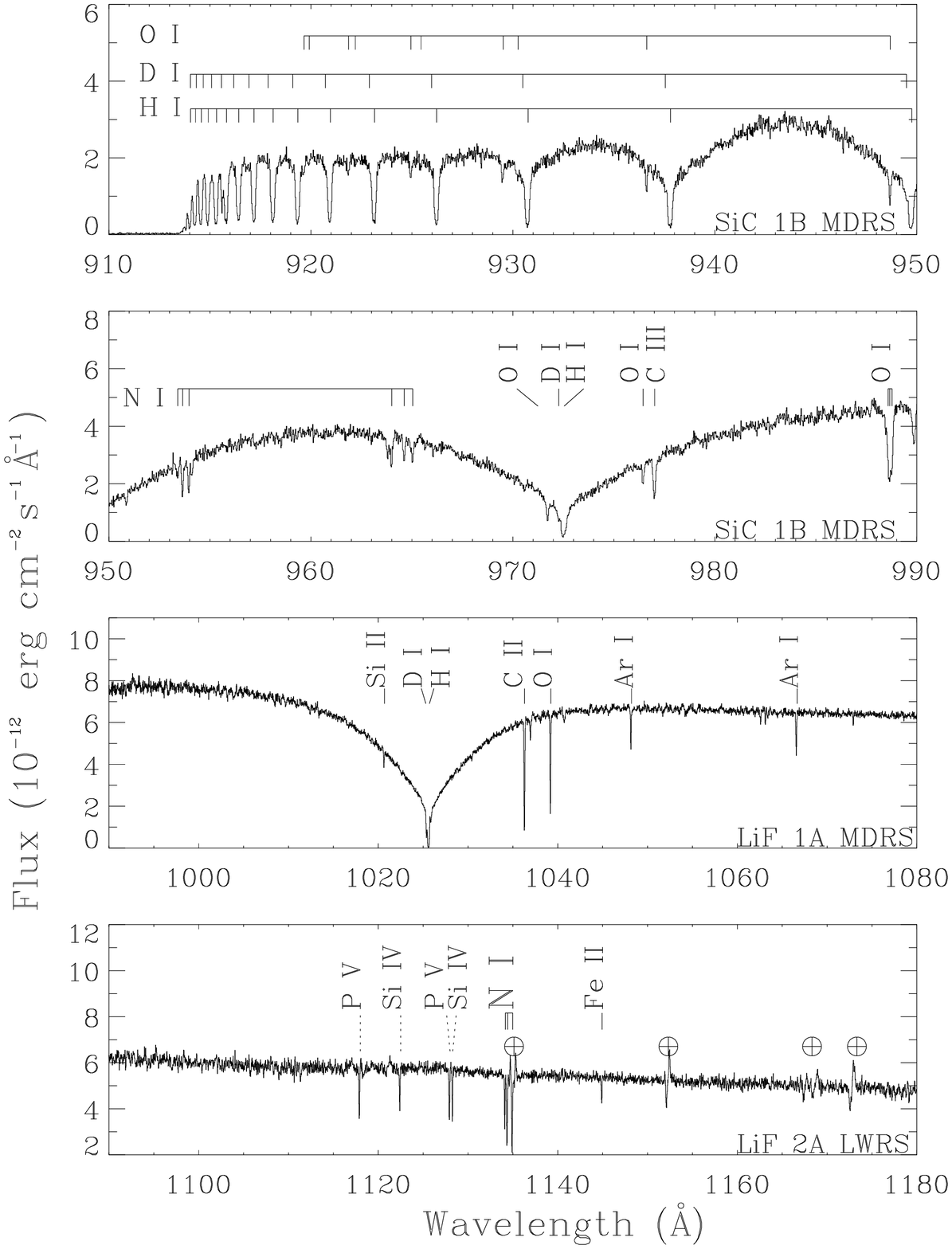}
\caption{\fuse~spectra of GD 246 with identification of interstellar lines. An annotation on each panel indicates which channels are displayed and if the data was obtained through the LWRS or MDRS apertures. Photospheric lines are marked with dotted lines and geocoronal emission is annotated with $\bigoplus$. Ar I $\lambda$1066 is blended with stellar Si IV. The data are binned by 4 for display purposes only. \label{gd246fusedata}}
\end{figure}

\begin{figure}
\epsscale{0.80}
\plotone{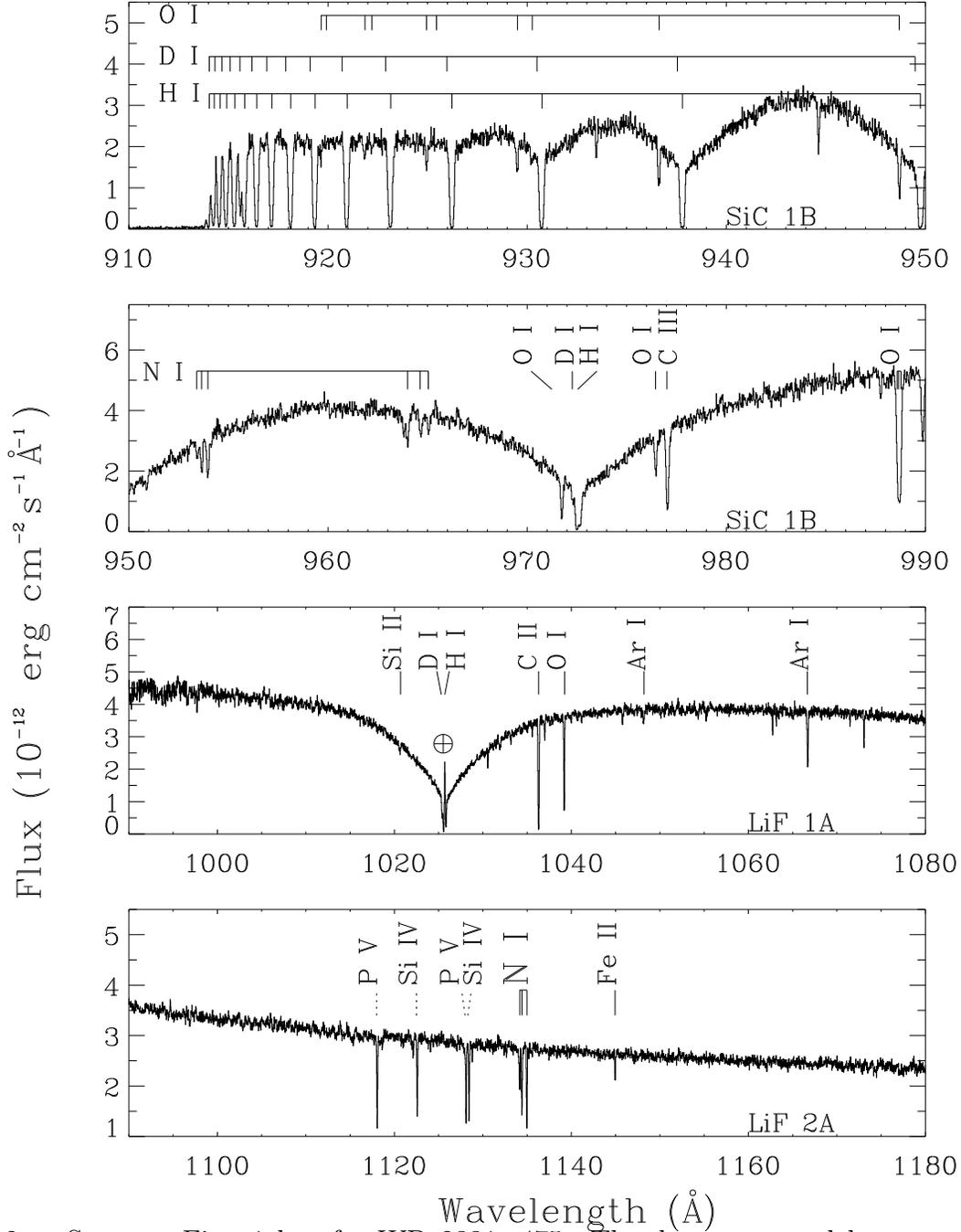}
\caption{Same as Fig. \ref{gd246fusedata} but for WD 2331$-$475. The data presented here are from the MDRS observation, P1044203. \label{wd2331fusedata}}
\end{figure}

\begin{figure}
\epsscale{1.00} %no epsscale here
\epsscale{0.80}
\plotone{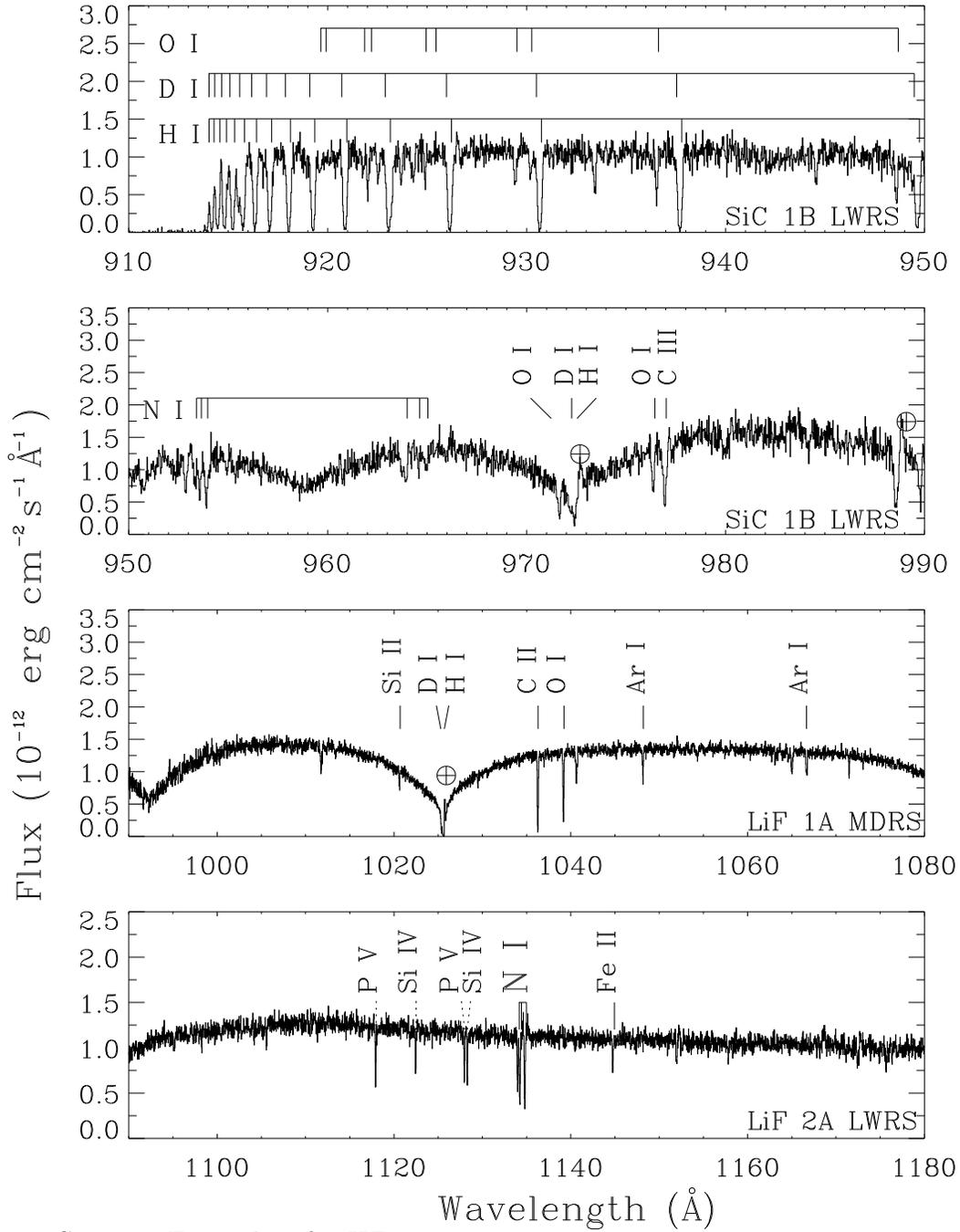}
\caption{Same as Fig. \ref{gd246fusedata} but for HZ 21.\label{hz21fusedata}}
\end{figure}

\begin{figure}
\epsscale{0.80}
\plotone{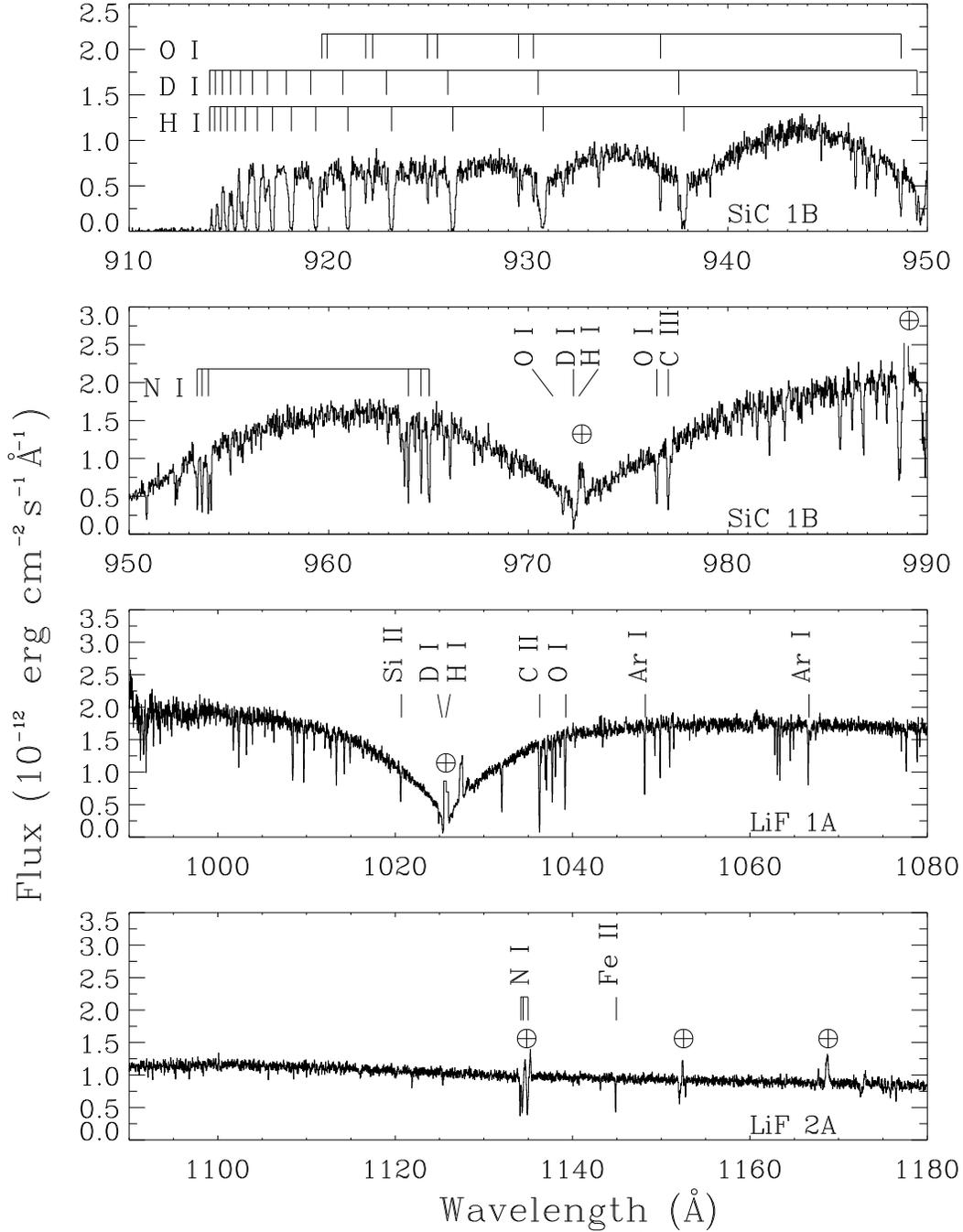}
\caption{Same as Fig. \ref{gd246fusedata} but for Lan 23. Many lines of molecular hydrogen (not marked) are also seen along this line of sight, for example in the range 1000~--~1020~\AA. \label{lan23fusedata}}
\end{figure}

Table \ref{fuse_obs} summarizes the \fuse~observations of the four white dwarfs studied in this work. The \fuse~spectrum of the hot DA white dwarf GD 246 is shown in Figure \ref{gd246fusedata}. Observations were obtained through both the large (LWRS, $30\mfarcs\times30\mfarcs$) and the medium sized apertures (MDRS, $4\mfarcs\times20\mfarcs$) in histogram mode (HIST, in which a two-dimensional spectral image is accumulated and downlinked). The spectrum of WD 2331$-$475 is shown in Figure \ref{wd2331fusedata}. Two observations were obtained through the LWRS aperture and one through the MDRS aperture. All of them were obtained in HIST mode. During one of the LWRS observations (P1044202) one detector (side 2) was not available resulting in the loss of data from the LiF 2A, LiF 2B, SiC 2A, and SiC 2B channels. The spectrum of HZ 21 is shown in Figure \ref{hz21fusedata}. The data were obtained through the LWRS and MDRS apertures, in time-tag mode (TTAG, in which the position and arrival time of each photon is recorded). The \fuse~spectrum of the hot DA white dwarf Lan 23 is displayed in Figure \ref{lan23fusedata}.
The data were obtained through the LWRS aperture in TTAG mode.

For the four stars the two-dimensional \fuse~spectra are reduced using the CalFUSE pipeline (version 2.0.5, Dixon, Kruk, \& Murphy, in preparation\footnote{The CalFUSE pipeline reference guide is available at http://fuse.pha.jhu.edu/analysis/pipeline\_reference.html}). The processing includes data screening for low quality or unreliable data, thermal drift correction, geometric distortion correction,
heliocentric velocity correction, dead time correction, wavelength calibration, detection and removal of event bursts, background subtraction, and astigmatism correction.
The spectra are aligned by cross-correlating the individual exposures over a short wavelength range that contains prominent spectral features and then coadded by weighting each exposure by its exposure time. All the spectra are binned to three pixel samples, or 20 m\AA, for analysis (the line spread function, LSF, is about 11 pixels or $\sim$ 70 m\AA~wide).

\subsection{STIS Observations of GD 246 and Data Processing}

Two \hst/STIS observations of GD 246 that include H I \lya~are available at the Multimission Archive at the Space Telescope Science Institute; one each using the E140H and E140M gratings.

Although the E140H grating has better resolution than the E140M grating it also suffers from more complications in the \lya~region. \lya~spans more orders in the E140H grating and the relative flux calibration of adjacent orders is not as good as the one for the E140M grating. Because the saturated H I \lya~line is so broad, the higher resolution of the E140H grating is not required. Therefore we make use of the E140M data for deriving the H I column density. We use the E140H data to study the abundance of metals such as C II*, Si II, and S II and to determine the velocity structure of the gas along the line of sight.

The STIS echelle-mode observations of GD 246 are summarized in Table \ref{STIS_obs}. The E140M (S/N $\sim$ 15, FWHM = 1.3 pix) and E140H (S/N $\sim$ 10 -- 25, FWHM = 1.2 pix) 
data are reduced using the standard STSDAS pipeline within IRAF. The CALSTIS pipeline (version 2.3) is used to produce the two-dimensional spectrum of the FUV-MAMA and then the scattered-light removal is performed 
using different algorithms for the different echelle gratings. For the E140M data the \citet{2000AAS...197.1202L} algorithm is used to estimate and remove the scattered light from the data in the pipeline procedure. After extracting the spectrum, two echelle spectral orders in the regions adjacent to \lya~are combined using a weighted averaging scheme where the orders overlap. 
For the E140H data the scattered light is removed with the procedure of \citet{2000AJ....119.2481H}.

\section{Analysis Methodology} % OK OK

Whenever possible we derive the column densities by more than one method. We use profile fitting (PF), curve of growth (COG) and apparent optical depth (AOD) methods. Table \ref{atomicdata} indicates which methods are used, for each transition, in the four stars. Below we discuss the techniques used in more detail.

\subsection{Profile Fitting}

We use the profile fitting code {\bf Owens.f} developed by Martin Lemoine and the French \fuse~Team to derive the column densities presented in this study. In this approach each interstellar absorption line is represented by the convolution of a theoretical Voigt absorption profile with the instrument line spread function (LSF), taken to be a single Gaussian in this work. An iterative procedure which minimizes the sum of the squared differences between model profiles and the data is used to determine the most likely column densities, $N$, radial velocities, and velocity dispersions of each component. The stellar continua in the vicinity of the absorption lines being studied are normalized by Chebyshev polynomials of low order.
Unsaturated lines from different elements in different spectral regions (designated ``windows'') are fit simultaneously to give a single solution for the temperature, non-thermal velocity, and radial velocity of each cloud (we define as unsaturated, those lines for which the absorption profile, before convolution with the instrumental LSF, has a residual intensity in the line core larger than 0.1). The software also allows us to use a wavelength dependent LSF and relative wavelength shifts between different absorption lines as free parameters. These last two features are particularly convenient when fitting \fuse~data, for which the LSF, while poorly constrained, is known to vary with wavelength. There are also small wavelength shifts between and within channels which are accommodated by this approach. For more on profile fitting with this code see \citet{2002ApJS..140..103H} and \citet{2002ApJS..140...67L}

Two of us (C. Oliveira and G. H\'{e}brard) performed the analyses for GD 246, WD 2331$-$475, HZ 21, and Lan 23 independently. In both of the analyses a Gaussian LSF is used. However, in one analysis (PF1) the FWHM of the LSF for each spectral window is allowed to vary independently during the fitting process, while in the other (PF2) it is not allowed to vary.
In PF2 there is an initial step in which all LSF's are free parameters of the fit. After a few trial fits the LSF's for the different windows converge and do not change significantly (these values are close to each other and close to what is known about the \fuse~LSF). At this point the LSFs are fixed at the newly found values, reducing the number of free parameters, for the remainder of the \chisq~minimization. We find typical LSF values of 8 -- 13 pixels (FHWM, 53.6 -- 87.1 m\AA) for the PF1 analyses; for the PF2 analyses we find a range of 9 -- 13 pixels for the LSF (FWHM, 60.3~--~87.1 m\AA). The background levels used for each line are those evaluated at the bottom of the closest Lyman line \citep[see][]{2002ApJS..140..103H}. Where possible, we compare the different channels and datasets to check for fixed-pattern noise in the regions of the absorption lines used in this study. When there is an inconsistency between channels and/or observations we remove that absorption line from our fit.

The error bars for each estimated column density are computed from an analysis of the \chisq~variation. The column density for which the error bars are being sought is fixed at a series of trial values, and for each trial value we compute the best fit and resulting \chisq~distribution where all the other parameters are allowed to vary freely. To define the confidence interval we use the distribution of \chisq. Scanning the column density in this way, we obtain the value of \chisq~as a function of the column density and derive the 1, 2, 3, 4, 5, 6, and 7$\sigma$~confidence levels using the standard $\Delta$\chisq~method. The 1, 2, 3, 4, 5, 6, and 7$\sigma$ error bars are divided by 1, 2, 3, 4, 5, 6, and 7, respectively. The average of these values, multiplied by two, is then used to produce the final 2$\sigma$~uncertainties reported here.
For more on the $\Delta$\chisq~method see \citet{2002ApJS..140..103H} or \citet{2002ApJS..140...67L}. We adopt the wavelengths and oscillator strengths for molecular hydrogen from \citet{1993A&AS..101..273A,1993A&AS..101..323A} and use \citet{1991ApJS...77..119M} with updates from Morton (1999, private communication, see Table \ref{atomicdata}) for the other species.

Both analyses produce similar results and the final values quoted here reflect the combined effort of these analyses. The results for all the four lines of sight are summarized in Table \ref{cols}~below, in \S~5.

%{\large $\Longrightarrow$ INSERT TABLE 4}

\subsection{Curve of Growth and Apparent Optical Depth}

We use the curve of growth (COG) and apparent optical depth (AOD) techniques to estimate the column densities of some species as a consistency check of the profile fitting results.
Not all species analyzed by profile fitting are suitable for analysis with these two methods. 
In the cases where the continuum is difficult to place (such as D I), or where there is blending (such as the N I triplets) we do not construct a curve of growth, or use the apparent optical depth technique. Since in these two techniques each absorption line is studied individually, continuum placement in a ``problematic'' line is harder then when many different lines of different absorption strengths and different continua are analyzed simultaneously as is the case of profile fitting. For species with only a small number of transitions available (such as P II and Ar I) we do not construct a curve of growth but we are able to use the apparent optical depth method.

In the COG technique the measured equivalent widths are fit with a single component Gaussian curve of growth \citep{1978ppim.book.....S}. In this technique the Doppler parameter, $b$, and column density, $N$, are varied so as to minimize the \chisq~between the measured equivalent widths and a model curve of growth. The stellar continuum in the vicinity of each line is estimated using a low order Legendre polynomial fit to the data. 
Contributions from Poisson noise, uncorrected high frequency fixed pattern noise, uncertainties in the Legendre fit parameters, and systematic uncertainties in the continuum placement and velocity integration range are used to estimate the uncertainties in the measured equivalent widths \citep[see][for details on this procedure]{1992ApJS...83..147S}. 
We measure the equivalent widths of a particular transition in all segments where that wavelength is covered. All the measurements from the different segments are compared; those that differ by more than 2$\sigma$~are excluded from our analysis (the differences are probably due to fixed pattern noise).

In the AOD technique the column density is determined by directly integrating the apparent column density profile, $N_{a}(v)~=~3.768\EE{14}\tau_{a}(v)/[f\lambda($\AA$)]$, over the velocity range of the absorption profile \citep[see][]{1991ApJ...379..245S}. The continuum is placed in the same way as described above in the COG method. This technique only yields the true total column density when the absorption is weak ($\tau~\le$~1) or if partially saturated, the components of the lines are fully resolved.

All the lines used in PF, COG, and AOD analyses are shown in Table \ref{atomicdata} along with log~$f\lambda$~for each transition. For each star, P, C, and A denote transitions that are used in profile fitting, curve of growth, and apparent optical depth, respectively. Both the curve of growth and apparent optical depth techniques produce results which generally have higher uncertainties, but are accurate enough to use as consistency checks.
These checks show that the AOD and COG results are within the 2$\sigma$~uncertainties of the column densities determined using profile fitting. 

\section{ANALYSIS OF THE GD 246 LINE OF SIGHT}

GD 246 is a hot hydrogen-rich (DA) white dwarf with effective temperature and gravity $T_{eff}~\sim$ 53,000K and 
log~$g~\sim$~7.85. The \fuse~spectra of GD 246 shows a smooth stellar continuum against which the 
interstellar lines are cleanly seen. Apart from the broad stellar Lyman lines, only a few stellar lines are 
present (Si IV, P IV--V). No molecular hydrogen and O VI lines are detected along this line of sight.

\begin{figure}
\plotone{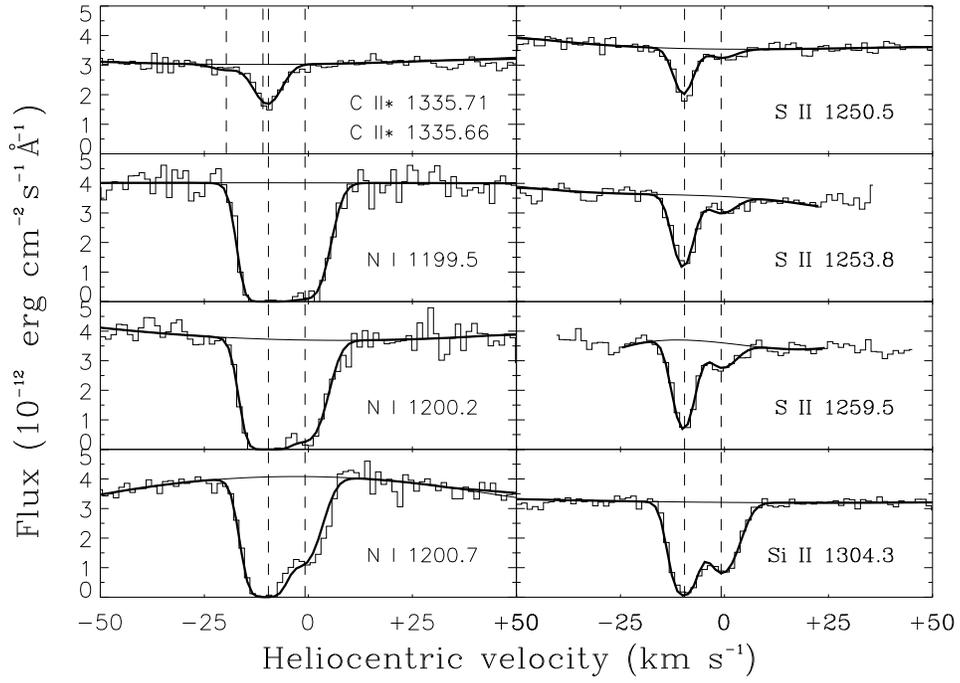}
\caption{STIS echelle spectra of GD 246, showing two absorption components. The two components at $v_{\odot}$~=~$-$9.8 and $-$0.6~km~s$^{-1}$ are marked by dashed lines. The weak component at $-$0.6~km~s$^{-1}$ is identified with the LIC in \S7.5. The bold line represents the fit to the absorption lines. \label{stisdata}}
\end{figure}

High-resolution STIS echelle spectra of GD 246 show that there are at least two velocity components, separated by $\sim$~9 km s$^{-1}$, along this line of sight. These are seen in the N I, S II, and Si II absorption profiles (see Figure \ref{stisdata}). As discussed in \S~7.5, the weaker component likely probes the LIC.

The two profile fitting analyses of this line of sight are different in an additional way.
In one of the profile fitting analyses (PF1) of this line of sight we use only \fuse~data with a variable LSF; in the other profile fitting analyses (PF2) we use both \fuse~and STIS data with fixed, but different LSFs, for the \fuse~and STIS spectral windows.
Both analyses are described in more detail below.

%In the profile fitting of this line of sight we fit together D I, N I, O I, C II*, S II, Fe II, Si II, Ar I and P II {\bf check order}.
%{\bf should describe the two different analysis in more detail, what species fit in each one.}

\subsection{Profile Fits}

In PF1 we fit a one component model to D I, C II*, N I, O I, Si II, P II, Ar I, and Fe II using \fuse~data only. The FHWM of the LSF is a free parameter of the fit for each spectral window being fit.

For PF2 we use high resolution STIS E140H spectra (R~$\sim$~114,000) of GD 246, covering the wavelength region 1170~\AA~--~1372~\AA, to constrain the velocity structure along the line of sight as well as to measure the column density of species with no transitions in the \fuse~bandpass (such as S II). Species which have transitions in the \fuse~and STIS bandpasses are fit simultaneously in both datasets. For the STIS data we use a single-Gaussian LSF with a FWHM of 2 pixels, for all the spectral windows. In this analysis we fit one absorption component to D I, C II*, N I, O I, Si II, S II, and Fe II, at the best-fit velocity of $v_{\odot}$~=~$-$9.76 km s$^{-1}$. For N I, Si II, and S II we use two components; component 1 at the best-fit velocity of $v_{\odot}$~=~$-$9.76 km s$^{-1}$ and component 2 at the best-fit velocity of $v_{\odot}$~=~$-$0.63 km s$^{-1}$.

The decision to distribute the second set of ions among 2 components is based on the following arguments. From an independent analysis of the STIS data alone, where we fit two components to N I, Si II, and S II, we determined that component 1 contains most of the material seen along this line of sight and that the Doppler parameters of the two components are similar. The ratio of column densities of component 1 to component 2 are N(N I)$_{1}$/N (N I)$_{2}$~=~5.9, N(S II)$_{1}$/N(S II)$_{2}$~=~4.6, and N(Si II)$_{1}$/N (Si II)$_{2}$~=~2.3.  
For species for which there is no STIS data, a velocity separation of $\sim$ 9 km s$^{-1}$ between the two components is not enough for profiles at the \fuse~resolution to produce meaningful results about the relative strength of the two different components. Since the physical parameters of the components seem to be similar we would expect component 1, the stronger one, to saturate first. However, we chose lines in the \fuse~data that appear unsaturated so that we did not have to worry about hidden saturation effects that would bias our column density determinations. For these reasons we choose to fit two components for the species for which we have STIS echelle data and one component for the species with no STIS data. In the end, the results of the two different approaches, PF1 and PF2, are consistent at the 2$\sigma$ level, with most consistent at the 1$\sigma$ level.

\begin{figure}
\epsscale{0.80}
\plotone{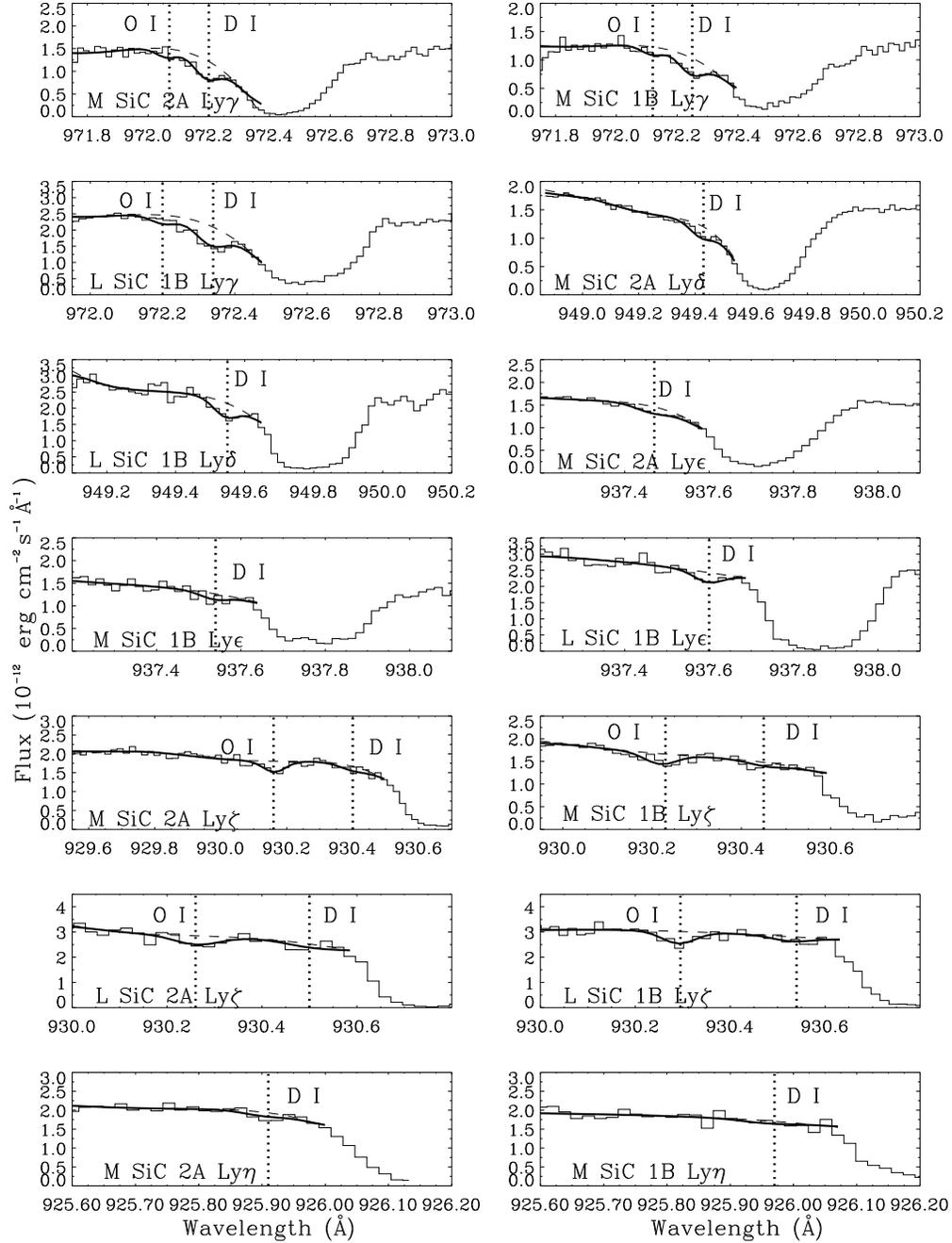}
\caption{GD 246 - Fits to the D I lines used in the analysis. L and M stand for LWRS and MDRS data, respectively. D I \lyb~was not used in the analysis due to contamination by geocoronal emission. Dashed vertical lines mark the centroids of the fitted D I and O I absorption lines. \label{gd246_di}}
\end{figure}

Because of the better resolution and signal-to-noise ratio of the STIS E140H data, we have tighter constraints on the column densities of C II*, S II and, Si II, in the model where STIS and \fuse~data are fitted simultaneously. For this reason the column densities and error bars reported for these species (Table~\ref{cols}) are the ones obtained with PF2. Profile fits for D I (PF2) are presented in Figure \ref{gd246_di} and the combined results are presented in Table \ref{cols}.

%{\large $\Longrightarrow$ INSERT FIGURE 5}

\subsection{Determination of the  H I Column Density Toward GD~246}  % OK OK OK

The H I column density along this sightline is determined by fitting the strong damping wings of interstellar \lya~using a newly computed model atmosphere to account for the stellar continuum. We consider the influence on the \lya~profile of several stellar models with different effective temperatures and gravities. \hst/STIS E140M ($\lambda$/$\Delta\lambda$~$\equiv$~R~$\approx$~45,800) archival data are used in this analysis.

%For the E140H data the 
%scattered light was removed with the Howk \& Sembach procedure. After extracting the spectrum, two echelle spectral orders in the regions adjacent to Ly$_{\alpha}$ were combined using a weighted averaging scheme where the orders overlap.
%To determine the interstellar H I column density we used archival observations taken with the Space Telescope Imaging Spectrograph (STIS) on board the \hst.  
%explain why we used only E140M data when E140H data is also available
%{\bf mention that also fit two components, with fixed separation and free separation etc. and the results are consistent at the 1 sigma level.}

\subsubsection{Stellar Model} % OK OK OK 

Several investigators measured the atmospheric parameters of GD~246 by fitting the Balmer lines with the use of either LTE or non-LTE (NLTE) stellar atmospheres models. They obtained effective temperatures and gravities that range from 53,000 K $\leq T_{\rm{eff}} \le 60$,000 K and $7.7 \le \log g \le 8.0$ \citep[see, e.g.,][]{1997ApJ...488..375F,1997MNRAS.286..369M,1997ApJ...480..714V,1998MNRAS.299..520B,1999ApJ...517..399N}. For our analysis of the stellar continuum in the vicinity of the \ion{H}{1} Ly$\alpha$ profile, and knowing that NLTE effects are important in the core of Ly$\alpha$, we adopted the atmospheric parameters determined by \citet{1999ApJ...517..399N} who based their results on pure-hydrogen NLTE models.  They obtained $T_{\rm{eff}} = 53$,$088 \pm 968$ K and $\log g = 7.85 \pm 0.07$ (1$\sigma$ error bars). \citet{1999ApJ...517..399N} pointed out that systematic uncertainties, which may result from the extraction of the spectrum, the flux calibration, or the normalization, are much larger than the statistical errors. To take into account these systematic errors, we considered the 1$\sigma$ errors on $T_{\rm{eff}}$ and $\log g$ to be 3,000 K and 0.15. The grid of models that we computed to estimate the stellar Ly$\alpha$ line includes these errors ($T_{\rm{eff}} = 50$,000 K, 53,000 K, and 56,000 K; $\log g = 7.70$, 7.85, and 8.0).

Observations in the EUV, FUV, and UV wavelength bands reveal that the atmosphere of GD~246 contains low abundances of elements heavier than hydrogen. For instance, \citet{2001ewwd.work...90C} and \citet{2001A&A...373..674W} analyzed the {\it International Ultraviolet Explorer} and \fuse~spectra of GD~246, and measured very low abundances of carbon, silicon, and phosphorus. \citet{1998A&A...329.1045W} analyzed the {\it Extreme Ultraviolet Explorer} spectrum of GD~246 with LTE metal-line blanketed models and concluded that the star had a lower metallicity than G191-B2B, which is the metal-rich white dwarf prototype. \citet{2000AAS...197.8305D} reported the detection of \ion{Fe}{6} absorption features in the CHANDRA LETG spectrum of GD~246. But the non-detection of \ion{Fe}{5} in the STIS spectrum of GD~246 puts an upper limit on the iron abundance of $\log({\rm{Fe/H}}) \le -6.0$. We expect that these small traces of heavy elements do not modify the physical structure of the atmosphere of the star, and consequently, do not influence the Ly$\alpha$ line profile. Therefore, we computed pure-hydrogen NLTE stellar atmospheres models to describe the stellar Ly$\alpha$ line.

\begin{figure}
\epsscale{0.75}
\rotatebox{90}{
\plotone{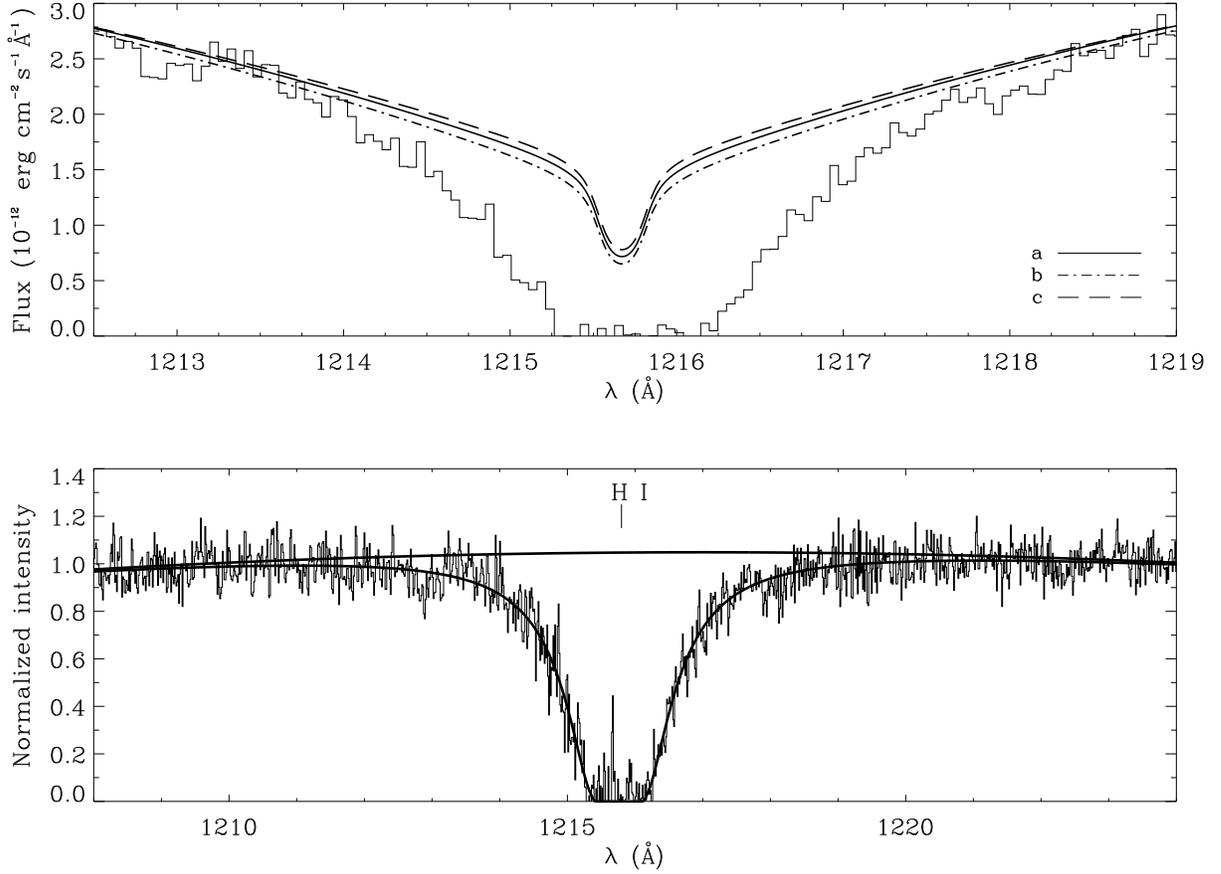}
}
\caption{Top panel: Interstellar H I \lya~toward GD 246 (data binned by 4 for display purposes only). The histogram shows the coadded STIS E140M observations of GD 246. Model $a$ is the best fit stellar model with \teff~=~53,000 K and log $g$~=~7.85 (see text for discussion). Model $b$ was computed with \teff~=~50,000 K and log $g$~=~7.7, and model $c$ with \teff~=~56,000 K and log $g$~=~8.0. Bottom panel: Fit to the H I \lya~profile after division by the best fit stellar model (model $a$, see text) and with the continuum modeled by a 2$^{nd}$ order polynomial. \label{lyman}}
\end{figure}

Determining the \ion{H}{1} column density with different stellar models places more credible error bars on N(\ion{H}{1}) than using a single stellar model. The statistical uncertainties are derived by using the best stellar model, and the systematic uncertainties associated to the stellar models are estimated by considering the most extreme stellar parameters, i.e., those parameters that yield the strongest and weakest stellar Ly$\alpha$. Figure \ref{lyman} illustrates such stellar Ly$\alpha$ profiles. The figure shows that the best stellar model (model a; $T_{\rm{eff}} = 53$,000 K and $\log g = 7.85$) produces a Ly$\alpha$ line profile that is between the strongest Ly$\alpha$ line (model b; $T_{\rm{eff}} = 50$,000 K and $\log g = 7.70$) and the weakest one (model c; $T_{\rm{eff}} = 56$,000 K and $\log g = 8.00$).

\subsubsection{\lya~Profile Fitting}  % OK OK OK

 \citet{2001A&A...373..674W} determined that the photospheric lines of this star are separated by $v$~=~$-$6 km~s$^{-1}$ from the interstellar lines. To take into account the relative shift between interstellar and stellar lines we align the model and the data in three different ways; displacing the model by $-$10, 0, and +10 km s$^{-1}$ relative to the data. By comparing the stellar Ly$\alpha$ model to the data, we took into consideration the radial velocity of GD~246, which includes its velocity with respect to the Sun and its gravitational redshift. Because of the large breadth of the interstellar H I absorption, displacing the model and the data by 10 km s$^{-1}$ does not lead to any difference in the measured H I column density from our fits. 
For each stellar model computed, the E140M spectrum of H I \lya~is normalized by the stellar model prior to profile fitting (see Figure \ref{lyman}, bottom panel). In addition, a 2$^{nd}$ order polynomial is used to model the continuum after normalization by the model atmosphere, to take into account uncertainties in the instrument sensitivity. Using the best fit stellar model we perform two fits (see \S~3 for a description of the fitting process) of the H I \lya~profile using both a 2 pixel single-Gaussian LSF and the tabulated STIS LSF for the $0\farcs02\times0\farcs09$ slit. There are no differences in the results using the two LSFs.
We derive log N(H I)~=~19.11~$\pm$~0.02~$\pm$~0.03 (2$\sigma$) where the first error bars reflect the uncertainties associated with the different stellar models and the second error bars take into account the statistical uncertainties when using the best fit model. We find then, log N(H I)~=~19.11~$\pm$~0.05 (2$\sigma$), by taking a conservative approach to combine the two different uncertainties. Our H I column density is in close agreement with the value of log N(HI)~=~19.12~$\pm~^{0.02}_{0.10}$~(1$\sigma$) derived from \euve~data by \citet{1997MNRAS.286...58B} using stellar models incorporating heavy elements.

\citet{2002ApJS..140...67L} have expressed concern over the presence of weak H I components which could perturb or bias the measurement of N(H I), but which would not be detected in other species due to their weak column density. \citet{2002ApJ...571L.169V} reanalyzed the Capella line of sight and showed a significant increase in the uncertainty due to this effect. However, they also concluded that for warm gas with column densities of H I above 10$^{19}$~cm$^{-2}$ the H I column density evaluation is more credible than for absorbers with lower column densities. We have tried to investigate the possible impact of additional weak, hot components on the determination of N(H I). In order to do so, we added one or two additional components in H I
only and performed the fits to \lya~and to \lyb. For \lyb~we used the blue wing of night only data, since the geocoronal emission in this region distorts the shape of the red wing of the profile. We were not able to find weak hot components of the type described by \citet{2002ApJS..140...67L} for which both \lya~and \lyb~profiles agreed. As pointed out by \citet{2002ApJS..140...67L} these weak hot components are very difficult to find and it is possible that we might have missed them, thus underestimating the uncertainties quoted for N(H I). However, we think this is unlikely given the conclusions of \citet{2002ApJ...571L.169V}.

\section{ANALYSIS OF THE WD 2331$-$475, HZ 21, AND Lan 23 LINES OF SIGHT}

WD 2331$-$475 is a hot DA white dwarf with \teff~= 51,800 K and log $g$ ~=~7.79 \citep{1997ApJ...480..714V}. A few stellar lines of Si IV, P VI--V, S IV--V are present in the data. There are no \hst~observations of this star and the quality of the \iue~data is not high enough for a reliable H I measurement.

HZ 21 is a helium rich (DO) white dwarf showing only a few stellar lines of N IV, P V, Si IV and S VI in the \fuse~data. Figure \ref{hz21fusedata} shows the broad stellar He II stellar absorption lines due to transitions from the energy level $n$ = 2.
 With effective temperature \teff~=~53,000 K and gravity log $g$ ~=~7.8 \citep{1996A&A...314..217D} this low metallicity star provides a smooth stellar continuum against which the interstellar lines are clearly seen. No molecular hydrogen is detected along this line of sight. There is probably more than one velocity component along this line of sight, considering the large distance to this star, $d$~=~115 pc. Unfortunately no \hst~observations of this star with sufficient resolution to resolve the velocity structure exist. In addition, this star has not been observed by \euve~and the quality of the \iue~data is not high enough for a reliable H I measurement.

Lan 23 is a DA white dwarf with an effective temperature of \teff~=~59,360 K and log $g$ ~=~7.84 \citep{1997ApJ...480..714V}. Stellar lines from the ground states of O VI, S VI, and from an excited state of C III are seen in the \fuse~spectra of this star. 
This star has not been observed by \hst, so we do not have information about the velocity structure along this line of sight. Considering that this star is at a distance of $d$~=~122~pc, several absorption components would be expected.  \citet{1999A&A...346..969W} have used \euve~data to measure the interstellar H I column density along this line of sight. Because of the high interstellar column density of H I and also some unknown opacity in the white dwarf itself, the \euve~flux is very small, which implies a large uncertainty in the determination of the interstellar H I column density. They find log N(H I)~=~19.89~$\pm~^{0.31}_{0.04}$, where the error bars quoted include uncertainties in the photospheric composition. These authors do not mention the confidence level corresponding to the quoted error bars, so we assume these are approximately 1$\sigma$ uncertainties, which we use to calculate the 2$\sigma$ uncertainties quoted in Table \ref{ratios}.

\subsection{Column Density Determinations}

\subsubsection{WD 2331$-$475}

\begin{figure}
\epsscale{0.7}
\plotone{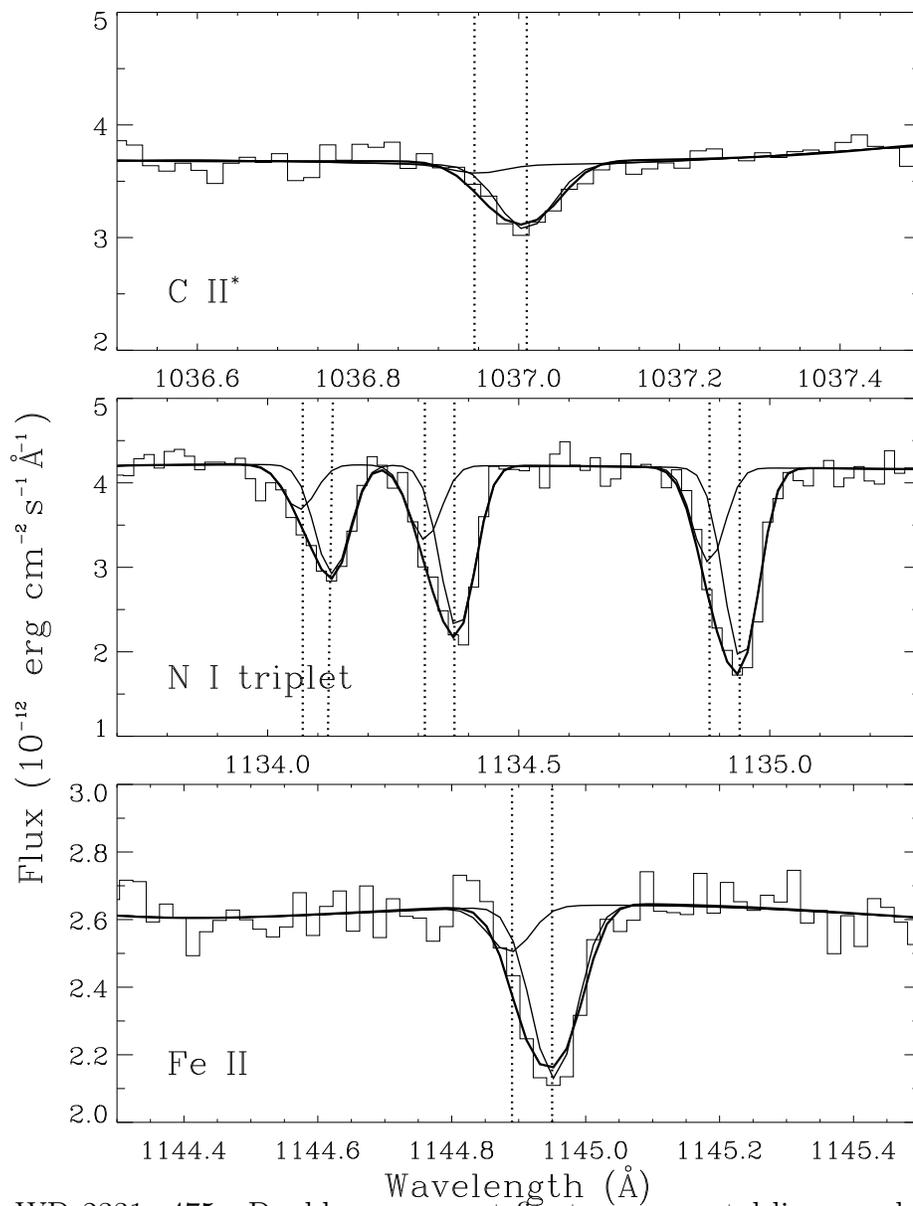}
\caption{WD 2331$-$475 - Double component fits to some metal lines used in the analysis. In this case the use of a single component showed significant differences between the results from the profile fitting and curve of growth analysis. In some cases, the discrepancy between the single component fit and the data was apparent upon inspection of the blue side of the absorption line. Dashed vertical lines mark the centroids of the two components for C II*, N I, and Fe II. \label{WD2331_2c}}
\end{figure}

\begin{figure}
\epsscale{0.85}
\plotone{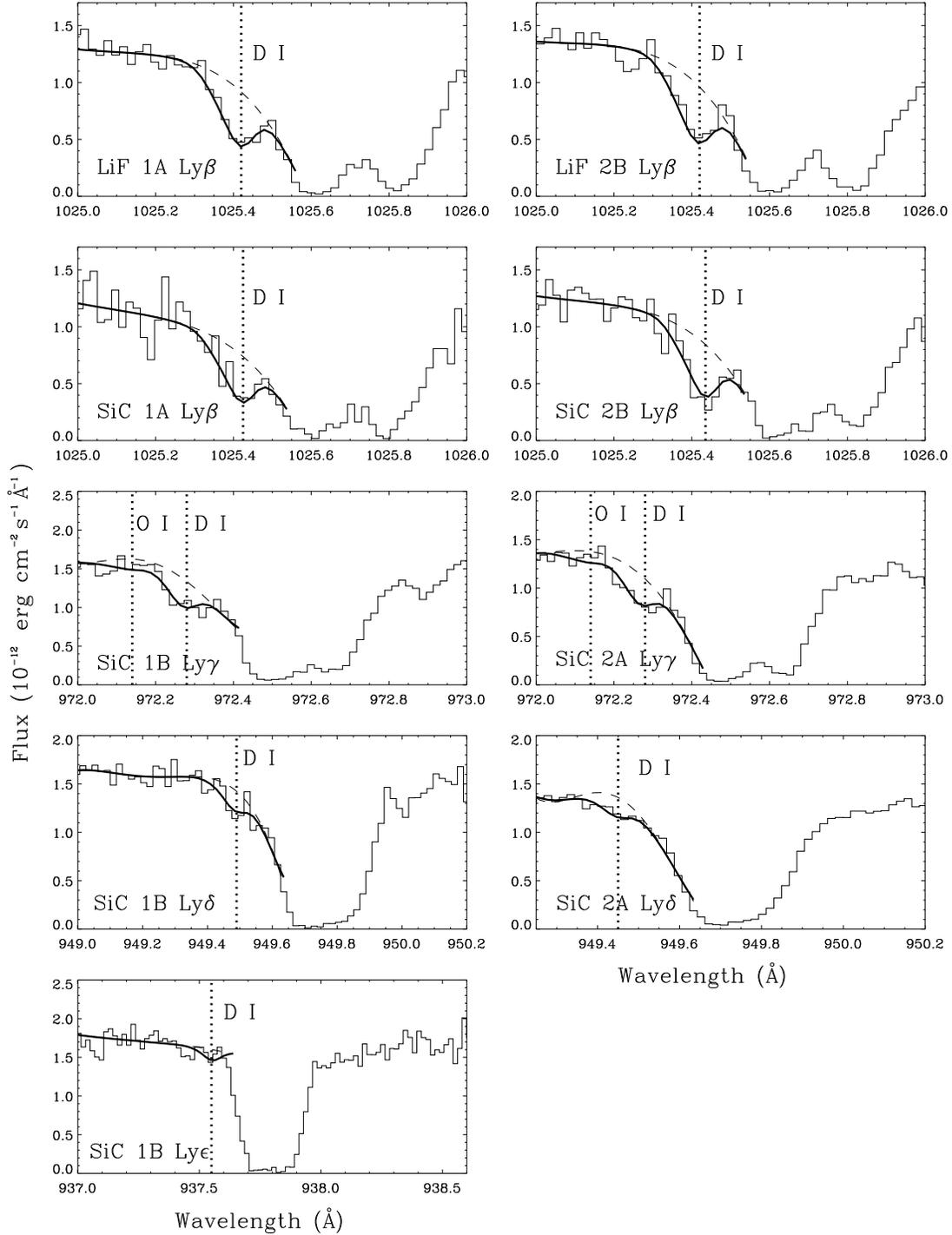}
\caption{WD 2331$-$475 - Fits to the D I lines used in the analysis, all from the P1044203 observation. For D I \lyb~only night data was used, in order to reduce the geocoronal emission. Some emission is still present between 972.5 and 972.7 \AA~and 1025.6 and 1025.8 \AA. Vertical dashed lines mark the centroids of the fitted D I and O I absorption lines.  \label{wd2331_di}}
\end{figure}

For this line of sight we measure the column densities of D I, C II*, N I, O I, Si II, P II, Ar I, and Fe II, using profile fitting, COG and AOD. Both PF1 and PF2 discussed previously were used. We first attempted to fit the data with a single absorption component. This resulted in a large underestimation of the D I, N I, and O I column densities when compared with the COG and AOD results. A closer inspection of the data revealed that some of the stronger lines have a wing extending towards negative velocities. Profile fitting with two components resulted in column density values within 1$\sigma$~of the COG and AOD results. We determined that the data is better represented by two absorption components separated by $\sim$14~km s$^{-1}$, with similar Doppler parameters, $b$. Figure \ref{WD2331_2c} displays the two component fit to C II*, N I, and Fe II. This component structure is also used for D I, O I, Si II, P II, and Ar I, not shown in Figure \ref{WD2331_2c}. The fit to D I (using PF2) is shown in Figure \ref{wd2331_di}. The column densities adopted are given in Table \ref{cols}.

\subsubsection{HZ 21 and Lan 23}

For both of these sightlines our profile fitting analyses assume the presence of only one interstellar component.
We fit the species D I, N I, O I, Si II, P II, Ar I, and Fe II in HZ 21 and D I, N I, O I, and Fe II in Lan 23. For the Lan 23 line of sight we include an extra component containing H$_{2}$, described in more detail in \S5.2.

\begin{figure}
\epsscale{0.85}
\plotone{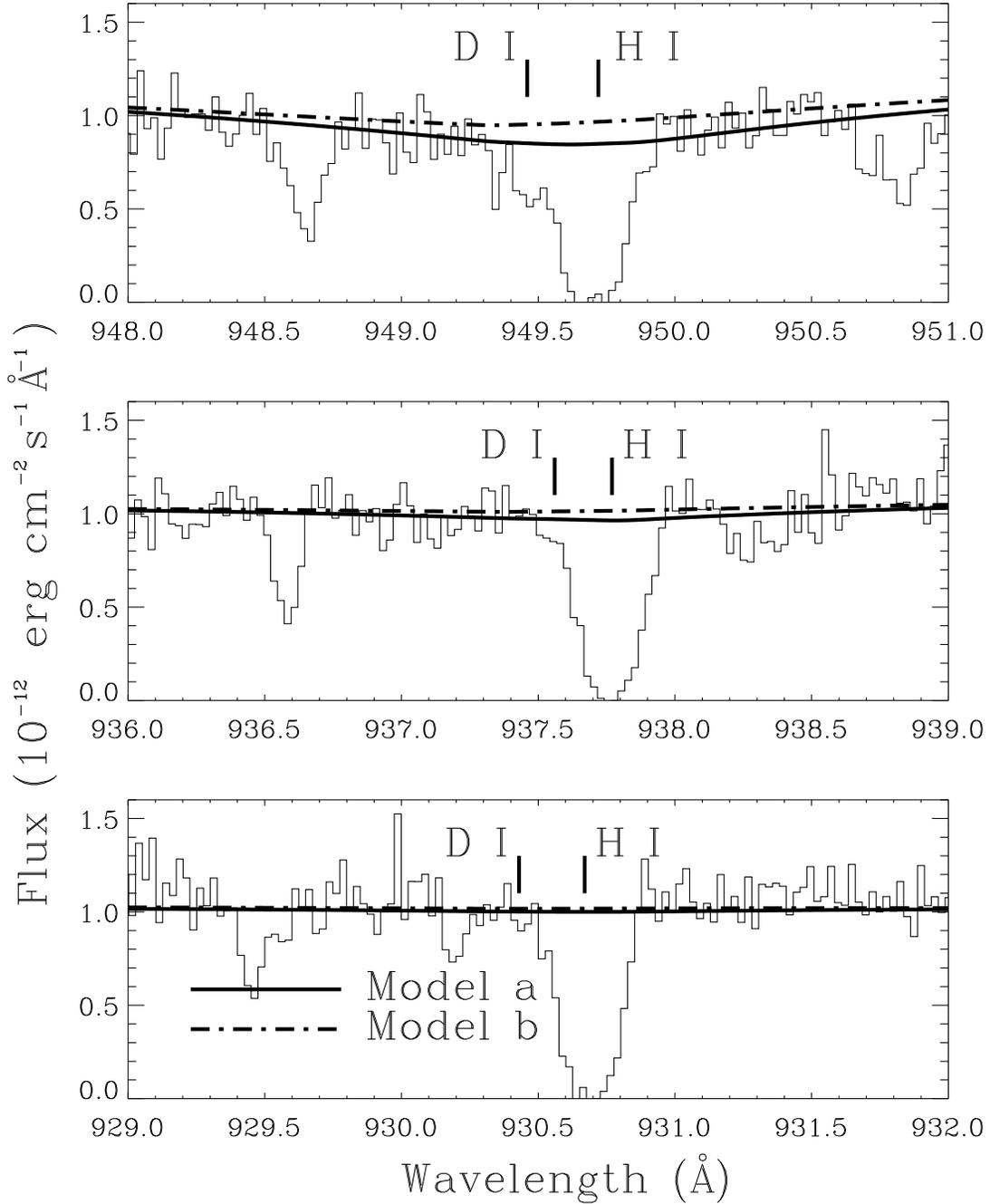}
\caption{Comparison between two models and the \fuse~data of the helium-rich white dwarf HZ~21 in the regions of the \ion{D}{1} $\lambda\lambda$930, 937, and 949 lines. The models are computed with $T_{\rm{eff}} = 53$,000 K, $\log g = 7.8$, and two stellar hydrogen abundances $\rm{N(H)/N(He)} = 0.1$ (model {\it a}) and  $\rm{N(H)/N(He)} = 0.01$ (model {\it b}). Thick vertical lines mark the positions of D I and H I. \label{hz21_model}}
\end{figure}

In the case of HZ 21, the uncertainties introduced by the presence of broad stellar He II absorption in the neighborhood of the D I lines should be negligible compared to our final error bars on the column density of D I. \citet{1996A&A...314..217D} determined an upper limit on the hydrogen to helium abundance of $\rm{H/He} < 0.1$ in the atmosphere of the helium-rich white dwarf HZ~21. To take into consideration the effects of the stellar helium abundance on the column density of \ion{D}{1}, we computed two stellar models with hydrogen to helium abundances of $\rm{H/He} = 0.1$ and 0.01. Figure \ref{hz21_model} displays two stellar models computed with $T_{\rm{eff}} = 53$,000 K and $\log g = 7.8$, but different H to He abundance ratios. Model $a$ has $\rm{H/He} = 0.1$ and model $b$ has $\rm{H/He} = 0.01$. In both models the He II stellar absorption is so broad that it does not affect the underlying D I absorption.

There are many oxygen transitions in the \fuse~bandpass covering a wide range of oscillator strengths. Unfortunately at the high column density of O I in Lan 23 and with the low signal-to-noise ratio of the data, it is difficult to find non-saturated O I lines that are suitable for our analysis. We find only two such lines, $\lambda$919.917 and $\lambda$974.070, which span more than 1 dex in log$f\lambda$. This is the source of the large error bars for the O I column density along this line of sight.

\begin{figure}
\epsscale{0.85}
\plotone{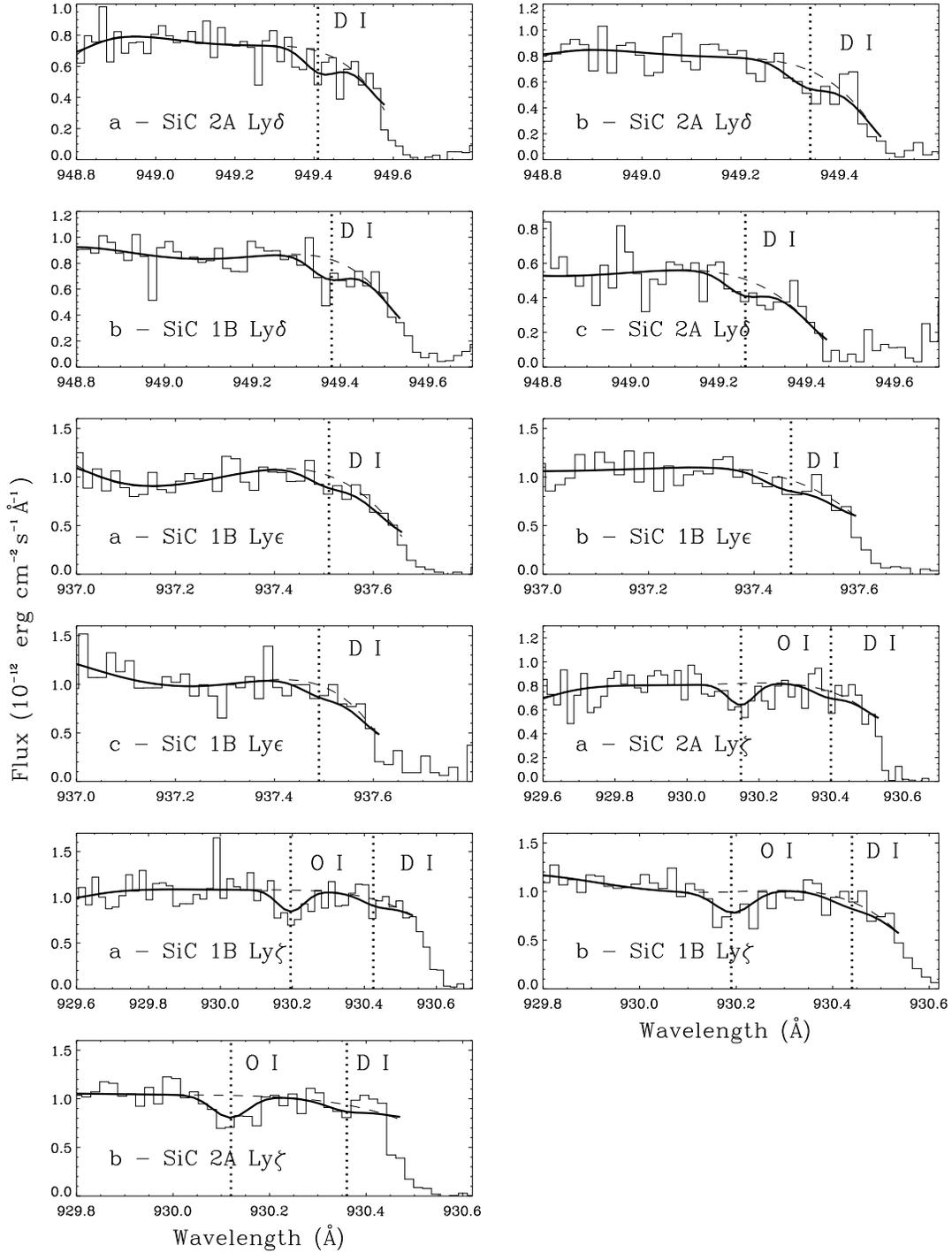}
\caption{HZ 21 - Fits to the D I lines used in the analysis. a, b, and c represent data from P204 MDRS, P204 LWRS, and M108 LWRS, respectively. Dashed vertical lines mark the centroids of the fitted D I and O I absorption lines. \label{hz21_di}}
\end{figure}

\begin{figure}
\epsscale{0.85}
\plotone{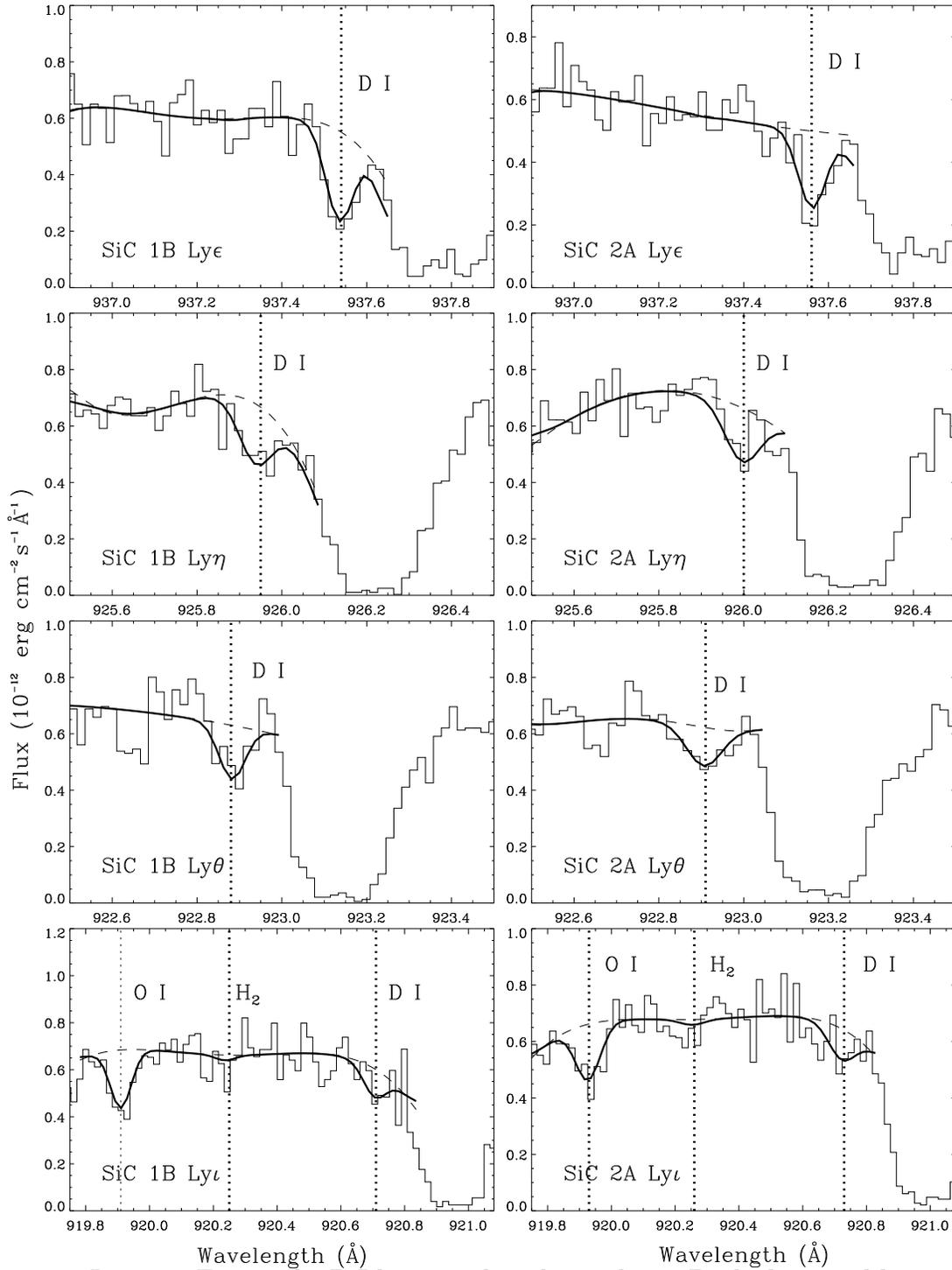}
\caption{Lan 23 - Fits to the D I lines used in the analysis. Dashed vertical lines mark the centroids of the fitted D I, O I, and H$_{2}$ absorption lines.\label{lan23_di}}
\end{figure}

Fits for D I (using PF2) are presented in Figures \ref{hz21_di} (HZ 21) and \ref{lan23_di} (Lan 23). The column densities found by combining the results of the two different analyses (PF1 and PF2), are given in Table \ref{cols}.

\subsection{Interstellar H$_{2}$ Toward Lan 23}

Absorption arising from H$_{2}$ rotational levels $J$ = 0 -- 3 is clearly seen in the \fuse~ spectra of Lan 23. Some H$_{2}$ lines are blended with D I and can be a problem for the measurement of the D I column density. We determine the column density of each $J$ level by fitting a single component curve of growth to lines from $J$~=~0 to 3, simultaneously, assuming the same $b$~value for all $J$ levels. From the curve of growth we obtain a Doppler parameter, $b$~=~4.1~$\pm~^{0.8}_{0.6}$ km~s$^{-1}$ (2$\sigma$). The $b$ obtained from the curve of growth is then used as a constraint in the profile fitting analysis of H$_{2}$, described in the paragraph below.

We also determine the H$_{2}$ column density by including unsaturated H$_{2}$ $J$ = 0~--~3 lines in our profile fitting analysis of this line of sight. The column density for each rotational level is determined independently, with no assumption about the rotational temperature of the gas. To allow for the possibility that this small amount of molecular hydrogen does not reside in the same gas as the D I, the velocity centroid of the H$_{2}$ is free to assume a different value from that of the D I.

Column densities obtained with curve of growth and profile fitting are displayed in Table \ref{h2cols}. The total H$_{2}$ column density along this line of sight is of the order log N(H$_{2}$)~$\sim$~15. The derived excitation temperatures are $T_{01}$ $\approx$ 291~K, $T_{12}$ $\approx$ 324~K, and $T_{23}$ $\approx$ 240~K, however, formal relative contributions of UV pumping versus collisional excitation have not been explored.

%{\large $\Longrightarrow$ INSERT TABLE 6}

%We find H$_{2}$($J$~=~0)~=~14.24$^{+0.05}_{-0.05}$, H$_{2}$($J$~=~1)~=~14.94$^{+0.06}_{-0.07}$, H$_{2}$($J$~=~2)~=~14.23$^{+0.06}_{-0.06}$, and H$_{2}$($J$~=~3)~=~13.94$^{+0.06}_{-0.07}$ (all errors are 2$\sigma$).

\section{SYSTEMATIC EFFECTS}

There are several systematic effects in the \fuse~data which can contribute to uncertainties in the estimated column densities and that are not directly included in our determination of the error bars.
Below we discuss the most important systematic effects in detail, how they can affect our data, and how we address these problems in our profile fitting analysis. For additional discussion of systematic effects in D/H measurements such as this, we refer the reader to \citet{2002ApJS..140..103H,2002ApJS..140...19K,2002ApJS..140....3M}.

\subsection{Fixed-Pattern Noise}

Fixed-pattern noise associated with the detectors can distort profile shapes and alter the equivalent widths of the absorption lines. The four channel design of \fuse~provides at least two (sometimes four) independent spectra of most of the lines used in the analysis. By comparing data from multiple segments, multiple observations and multiple slits we are able to identify the absorption profiles that suffer from large (compared to the statistical noise) fixed-pattern noise artifacts and exclude them from our fits. However, small effects cannot be ruled out.

\subsection{Background and Line Spread Function}

Although a comprehensive study of the line spread functions for the \fuse~instrument is not yet available, preliminary studies indicate that it is composed of two (Gaussian) components. The narrow component has a FWHM of $\sim$ 15 -- 20 km s$^{-1}$ and the broad component contains 20 -- 30$\%$~of the total LSF area (with a large portion of this broad component falling under the narrow core of the LSF). This broad component is responsible for the residual light in the cores of saturated lines (e.g., the high-order Lyman lines in Figures 1--4); in addition there may be scattered light. We use a single Gaussian LSF in our analysis and adopt a small zero flux level shift to account for the residual light in lines close to the \lya~series of H I, where this effect is more visible. Because we are fitting only non-saturated lines, any residual errors in this correction for the tail of the LSF should not affect our derived column densities in a significant way. To the extent that this small amount of residual light is due to the LSF, it may produce small systematic changes in the derived column densities, but such effects are expected to be small compared to the derived uncertainties.

\subsection{Oscillator Strength Uncertainties}

While oscillator strengths ($f$-values) of the D I lines are well known, the same may not be true for other elements such as N I and O I. By fitting multiple lines of the same species at the same time we are able to remove the lines that stand out as a poor fit and average out the small errors of the other lines in the final result. A case where the uncertainty in an O I $f$-value might pose a problem is in the determination of the O I column density for the Lan 23 line of sight. For this star, only two non-saturated lines have enough signal-to-noise ratio to be useful in our analysis: $\lambda$974.07 (very weak, $f$ = 1.56$\EE{-5}$ and partially blended with H$_{2}$(J = 2)) and $\lambda$919.9 ($f$ = 1.76$\EE{-4}$).

In other analyses of O I where $\lambda$919.9 was used \citep[e.g.][]{2002ApJS..140..103H} this line did not stand out as having a large problem with its $f$-value. A test, where the O I column density was determined by fitting each line individually, showed a 20$\%$ dispersion around the value obtained with a fit to all the lines simultaneously \citep[see][]{2002ApJS..140..103H}. Since the error bars derived for N(O I) toward Lan 23 are large, it is unlikely that the uncertainty in the $f$-values of the O I lines used here play a large role in our measurement of N(O I).

We measure the column density of Ar I in the four lines of sight using only one transition of Ar I ($\lambda$1048.2198) because the other one ($\lambda$1066.6599) is blended with photospheric Si IV. Any uncertainties in the $f$-value of this transition will affect our column density of Ar I.

\subsection{Number of Interstellar Components Along the Line of Sight}

In \S~4.2.2 we discussed the potential effects of weak hot H I components on the determination of N(H I) for GD 246. Because the H I column density is greater than 10$^{19}$ cm$^{-2}$, such effects are probably small.
For all species a major difficulty associated with measuring column densities is the presence of unresolved overlapping absorption components along the line of sight. Small scale velocity structure has been seen in, amongst others, K I and Na I ultra-high resolution surveys, for relatively small path lengths \citep[see for example][]{2001ApJS..133..345W}. The problem arises if the physical conditions in the different absorption components are significantly different. In a mix of cold (narrow) and warm (broad) components, each with approximately the same column density, the cold component will become saturated before the warm component. Because the components are unresolved, the absorption profile resulting from the sum of the two components may appear to be unsaturated, thus leading to an incorrect estimate of the total column density. For GD 246 the STIS data shows no evidence of such material. Lan 23 has a low H$_{2}$ column density ($\sim~10^{15}$~cm$^{-2}$), and the other white dwarfs show no evidence of H$_2$ in their \fuse~spectra.
However, without ultra-high resolution data this systematic effect can never be completely ruled out.

%in what direction does this work, underestimate or overestimate the column density?

\subsection{Profile Contamination by Unknown Stellar Features}

Although improbable, some of the lines used in our analysis could be contaminated by unknown stellar features \citep[e.g.][]{2002ApJS..140...51S,2002ApJS..140...37F}.
Since the four white dwarfs studied here are all metal poor, we do not believe this is a significant concern. Photospheric Si IV is present in the spectra of the white dwarfs studied here and is blended with Ar I $\lambda$1066.66; this Ar I transition is not used in our analysis.

\subsection{Effect on Column Density Ratios}

We believe that the systematic effects discussed above are generally small. It is possible that these effects could be of importance for the column densities with small statistical uncertainties. However, it is unlikely that they will be of significance for the column density ratios discussed in \S~7.

\section{RESULTS AND DISCUSSION}

In section 7.1 we discuss the line of sight to the four stars in the context of the Local Bubble.
In \S~7.2 we compare the total column density ratios of D/H, O/H, and N/H for GD 246, WD 2331$-$475, HZ 21, and Lan 23 to values found for other lines of sight in the local ISM and discuss the implications of using O and N as tracers of H for these lines of sight. 
In \S~7.3 we combine the ratios derived here with previously published \fuse~ratios, deriving revised \fuse~mean ratios.
In \S~7.4 we discuss the mean D/O ratio inside the Local Bubble.
\S~7.5 presents a more detailed discussion of the GD 246 line of sight. 
%Sightline-averaged depletions and the electron density in the Local Bubble are derived using C II * and S II. 
Because we have information on the velocity structure along this line of sight we can separate the LIC and Local Bubble contributions to the total column densities and draw more detailed conclusions about the distribution of material along the line of sight.

\subsection{Local Bubble}

\citet{1999A&A...346..785S} have used absorption line studies of Na I to approximately map the contours of the Local Bubble. They found a cavity with a radius between 65 and 250 pc (depending on the direction) that is delineated by a sharp gradient in the neutral gas column density with increasing radius, a dense neutral gas ``wall''. \citet{1999A&A...346..785S} quote an equivalent width of 20~m\AA~of Na I $\lambda$5891.59 as corresponding to log N(H I)~$\sim$~19.3, which is used as a rough estimate of a plausible Local Bubble boundary.
According to their map of the Local Bubble, the GD 246 line of sight does not penetrate the wall, as it lies between the 5 m\AA~and 20 m\AA~Na I $\lambda$5889.95 contours. Although the uncertainty in the distance to GD 246 (and for the other three white dwarfs, for that matter) makes this result ambiguous, log N(H I)~=~19.11~$\pm$~0.05 implying that GD 246 is likely close to the wall of the Local Bubble, but does not penetrate the higher density regions beyond. The lines of sight to WD 2331$-$475 and HZ 21 do not penetrate the wall either; WD 2331$-$475 lies inside the 20 m\AA~contour and HZ 21 lies between the 5 m\AA~and 10 m\AA~contours. In addition, as shown in the next section, the measured O I column density toward both WD 2331$-$475 and HZ 21 implies an H I column density similar to that towards GD 246. The Lan 23 line of sight is located between the 20 m\AA~and 50 m\AA~Na I $\lambda$5889.95 contours, and has a high H I column density; it likely penetrates the wall of the Local Bubble.

\subsection{O/H, D/H, and N/H Ratios}

Table \ref{ratios} summarizes the column density ratios for several species along the four lines of sight studied in this work. 
 For GD 246 we find O I/ H I~=~(3.63~$\pm~^{0.77}_{0.67}$)$\EE{-4}$. This ratio is in agreement with the O I/H I ratio derived by \citet{1998ApJ...493..222M}, of O I/H I~=~(3.43~$\pm$~0.30)$\EE{-4}$~(using the $f$-value updated to $f$~=~1.16$\EE{-6}$~from \citet{1999ApJS..124..465W}.

Reliable H I measurements are not available for WD 2331$-$475, HZ 21, or Lan 23. However, the constancy of the O/H ratio discussed above enables us to derive an approximate column density for H I using O I. Using the \citet{1998ApJ...493..222M} interstellar O I abundance  and the O I column density along each line of sight, we estimate log N(H I)~=~18.94 for WD 2331$-$475, log N(H I)~=~19.20 for HZ 21, and log N(H I)~=~20.16 for Lan 23.

Using the N(H I) obtained this way and our measured N(D I) we obtain D I/H I~$\sim$~1.8$\EE{-5}$, 1.6$\EE{-5}$, and 1.2$\EE{-5}$, for WD 2331$-$475, HZ 21, and Lan 23, respectively. The uncertainties in the O I columns used to derive the H I column densities are large enough that the differences between the derived D/H ratios for these stars and the \citet{2002ApJS..140....3M} weighted mean of D/H~=~(1.52~$\pm$~0.15)$\EE{-5}$ are not significant. The D/H ratio for GD 246 is discussed below in \S~7.5.

Using the \citet{1997ApJ...490L.103M} interstellar abundance of nitrogen, N/H~=~(7.5~$\pm$~0.8)$\EE{-5}$, and our measured N I column density leads to estimated values of log N(H I)~=~18.87 for GD 246, log N(H I)~=~18.65 for WD 2331$-$475, log N(H I)~=~18.89 for HZ 21, and log N(H I)~=~19.82 for Lan 23.
The H I column density derived using the N I column density is in all four cases only $\sim$50$\%$~of the H I column density determined using the O I abundance, implying that the N I column densities are low. It is possible that the N along these lines of sight is $\sim$50$\%$~ionized. This effect has been seen for other sightlines as well \citep[see][]{2000ApJ...538L..81J,2002ApJS..140...81L,2002ApJS..140....3M}. For GD 246 we measure N I/H I~=~(4.37~$\pm~^{0.84}_{0.74}$)$\EE{-5}$ which is a factor of 1.7 times smaller than the \citet{1997ApJ...490L.103M} value, but is in close agreement with the weighted mean of the \fuse~sightlines N I/H I~=~(4.24~$\pm$~0.62)$\EE{-5}$ \citep[see][]{2002ApJS..140....3M}.

The ionization fraction of N I along a line of sight can be estimated directly by measuring the N II and N III column densities as both N II and N III have transitions in the \fuse~bandpass. However, for our sightlines the N II transitions available in the \fuse~bandpass are saturated, and N III cannot be determined since it is blended with interstellar Si II and sometimes with photospheric N III. \citet{2002ApJS..140..103H,2002ApJS..140...19K,2002ApJS..140...91W} have been able to measure the column density of N II (and sometimes N III) along three different lines of sight and found that in all cases it was larger than the column density of N I, providing a better agreement between (N/H)$_{total}$ within 100 pc and N I/H$_{total}$ for larger path lengths.
 
\subsection{Revised \fuse~Ratios}

In Table \ref{comparison} we combine the ratios derived here with the \citet{2002ApJS..140....3M} ratios for D I/H I, O I/H I, N I/H I, D I/N I, D I/O I, and O I/N I, and present the \citet{1997ApJ...490L.103M,1998ApJ...493..222M} N/H and O/H values for comparison. We present also the $\chi^{2}_{\nu}$ test for a single mean value. The values quoted in Table \ref{comparison} are the weighted means and uncertainty in the means, where we use the largest of the lower and upper error bars of each individual ratio to compute the weighted mean. {\it We use 1$\sigma$~error bars in this subsection} in order to make it easier to compare the different lines of sight among themselves. For Lan 23 values only the D I/N I value is used, due to the large error bars for the other ratios associated with this line of sight . By combining our ratios with the \citet{2002ApJS..140....3M} values, the weighted means increase by 0 to 5\%.

The mean D I/O I ratio is (4.06~$\pm$~0.17)$\EE{-2}$ with $\chi^{2}_{\nu}$~=~1.9 ($\nu$~=~9).
The mean D I/N I ratio now computed with ten lines of sight is (3.41~$\pm$~0.15)$\EE{-1}$ with $\chi^{2}_{\nu}$~=~2.2 ($\nu$~=~9).

\subsection{D/O Ratio Inside the Local Bubble}

\citet{2001IAP} and \citet{2002ApJS..140....3M} have found a nearly constant D/O ratio inside the Local Bubble, indicating that D/O traces the D/H ratio.  The weighted mean of the five \fuse~sightlines within the Local Bubble \citep{2002ApJS..140....3M} is (3.76~$\pm$~0.20)$\EE{-2}$~(1$\sigma$ in the mean). When we combine the D/O ratios derived here (Lan 23 value not included, as it is likely outside the Local Bubble) with the Local Bubble \fuse~results from Moos et al. (2002), we derive a weighted mean and uncertainty in the mean of D I/O I~=~(3.87~$\pm$~0.18)$\EE{-2}$~(1$\sigma$ in the mean) with $\chi^{2}_{\nu}$~=~0.96 ($\nu$~=~7), reinforcing the Moos et al. (2002) conclusion that the variability of the data is consistent with the uncertainties, i. e., the D I/O I ratio is constant within the Local Bubble. The D I/O I ratio inside the Local Bubble will be discussed in more detail by H\'{e}brard et al. (2002, in preparation).

\subsection{GD 246}

Absorption by at least two different components, separated by $\sim$9 km 
s$^{-1}$, is clearly seen, along the GD 246 sightline, in the high resolution STIS echelle data of N I, S II, and Si II (see Figure \ref{stisdata}). The stronger component (component 1),
seen at $-$9.76 km s$^{-1}$, contains 86$\%$, 82$\%$, and 79$\%$ of the total column densities of N I, S II, and Si II, respectively, along this line of sight. 
The velocity of component 2, $v_{2}$~=~$-$0.63 km s$^{-1}$, is consistent with the expected velocity 
of the Local Interstellar Cloud (LIC) along this line of sight \citep[+2.46~$\pm$~1.84 km s$^{-1}$;][]{1995A&A...304..461L} when the zero point velocity uncertainty in STIS E140H data is taken into account ($\Delta$$v$~$\sim$~0.7 -- 1.5 km s$^{-1}$). The discrepancy between the expected LIC velocity and the measured velocity along this sightline might be an indication that the LIC velocity field is more complicated than a single vector. A similar effect has been seen toward the Hyades Cluster by \citet{2001ApJ...551..413R}, who found $v_{\rm{predicted}}$(LIC) $-$ $v_{\rm{observed}}$(LIC) = 2.9 $\pm$ 0.7 km s$^{-1}$, similar to our value $v_{\rm{predicted}}$(LIC) $-$ $v_{\rm{observed}}$(LIC) $\sim$ 3.1 km s$^{-1}$. The eighteen Hyades stars studied by these authors are located in the sky not too far frm GD 246; for four of them (HD 27561, HD 27848, HD 28237, and HD 29225; $l~\sim$ 178 -- 184, $b~\sim$ $-$20 -- $-$25) they also find $v_{\rm{predicted}}$(LIC) $-$ $v_{\rm{observed}}$(LIC) $\sim$ 3.0 km s$^{-1}$. 

Sulfur should not be depleted onto dust grains, making it possible to estimate how neutral hydrogen would be distributed between the two components using S II as a proxy for H I. We derive log~N(H I)$_{2}$~=~18.39 using the S II column density for the LIC component (component 2), log~N(S II)$_{2}$~=~13.59. We derive a distance of 8 pc to the edge of the LIC along this direction \citep[assuming a LIC neutral hydrogen column density of $n_{H I}$~=~0.10 cm$^{-3}$;][]{2000ApJ...528..756L}. Both the H I column density and distance to the edge of the LIC derived here are greater than the values derived from the model of Redfield (2002, private communication), for this direction, of log N(H I)~=~17.93 and d~=~2.75 pc. Our values are closer to the maximum values derived for the LIC by \citet{2000ApJ...534..825R} of 18.32 and 6.8 pc, respectively. 
The discrepancy between our values and the LIC model values can be due either to the preliminary nature of their model or to the presence of an unresolved cloud close to the LIC velocity. While H I may not follow exactly the same distribution as S II it is still clear that most of the neutral gas along this line of sight is outside the LIC but within the Local Bubble boundaries.

We find D~I/H~I~=~(1.51~$\pm~^{0.39}_{0.33}$)$\EE{-5}$ for this line of sight (2$\sigma$~error bars). So far, only three lines of sight studied by \fuse~(HZ 43A, G191-B2B, and WD 1634$-$573) probing the Local Bubble have accurate D/H measurements \citep{2002ApJS..140...19K,2002ApJS..140...67L,2002ApJS..140...91W}. This is the fourth line of sight, with an accurate D/H measurement that probes the region mentioned above.

The D I/H I ratio for all the gas in this sightline is (1.51~$\pm~^{0.39}_{0.33}$)$\EE{-5}$. \citet{1998SSRv...84..285L} reports a mean value of (1.50~$\pm$~0.10)$\EE{-5}$ (1$\sigma$ in the mean) for 12 LIC sightlines. Because most of the gas along the sightline to GD 246 resides outside the LIC, we conclude that the D I/H I ratio for the gas outside, is also $\sim$1.5$\EE{-5}$, similar to the LIC value.

%The gas outside the LIC must also have an average deuterium abundance of D/H~$\sim$1.5$\EE{-5}$, if one takes into account the mean value, D/H~=~(1.50~$\pm$~0.10)$\EE{-5}$, of 12 LIC lines of sight reported by Linsky (1998). This implies D and H column densities of log N(D I)~=~14.19 and log N(H I)~=~19.02 for the material outside the LIC.

The electron density in component 1, which probes material outside the LIC, can be computed under the assumption that electron collisions populate the upper fine structure level ($J$~=~3/2) of the ground state of C II leading to C II* \citep[see][]{1993ApJ...409..299S}. 
Because all the C II lines in the \fuse~and STIS data are saturated, S II is used as a proxy for C II. This is a reasonable assumption; C II and S II have similar ionization potentials (24.38 eV for C II and 23.33 eV for S II) and should therefore trace the same type of gas.
The collisional equilibrium equation is
\begin{equation}
A_{21}n(C^{+*})=\gamma_{12}n_{e}n(C^{+})
\end{equation}
where $\gamma_{12}$ is the rate coefficient for excitation and A$_{21}$ is the spontaneous downward transition probability. For excitation of ions by electrons \citep{1978ppim.book.....S}:
\begin{equation}
\gamma_{12}=\frac{8.63\EE{-6}}{g_{1}T^{0.5}}\Omega_{12}~exp(-\frac{E_{12}}{kT})~cm^3~s^{-1}
\end{equation}
$E_{12}$~=~7.9$\EE{-3}$~eV, and we adopt $A_{21}$~=~2.29 \citep{1983IAUS..103..143M} and $\Omega_{12}$~=~2.90 \citep{1989agna.book.....O}. Assuming that conditions are uniform within the cloud we can replace particle density ratios by column density ratios and assuming that for warm low density clouds the depletions of S and C are 0 and $-$0.3 dex respectively, we find
\begin{equation}
\frac{N(C^{+*})}{N(S II)}=\frac{55n_{e}}{T^{0.5}}~.
\end{equation}
Using the S II column density we measured for component 1, log N(S II)$_{1}$~=~14.25~$\pm$~0.02, and the C II* column density, log N(C II*)~=~13.05~$\pm$~0.04, we find n$_{e}$~=~(0.1~$\pm$~0.01)(T/8000)$^{0.5}$ cm$^{-3}$ (all error bars are 2$\sigma$). For T~=~4000--8000 K this yields an electron density, $n_{e}$~=~0.06--0.11 cm$^{-3}$ for the material beyond the LIC,  which is similar to values found in the LIC by \citet{2000ApJ...528..756L}.

Thus, the D/H ratio and the electron density obtained in this study indicate that the properties of the cloud(s) beyond the LIC along the GD 246 line of sight are likely similar to those of the LIC.
 
The presence of ionized gas in this cloud will not affect the D/H ratio in this region, as D I and H I, with a similar ionization potential, are ionized at the same rate. 
% For GD 246, where ionized He was first detected in the local ISM (Vennes et al. 1993), Barstow et al. (1997) find a ionization fraction of helium, $f_{He}$~=~$N_{He II}/(N_{He I}+N_{He II})$, between 0.17 and 0.30.
%We can predict what the sightline averaged fraction of H II is, by using the approximation N(H${^+}$)~=~N(H$^{0}$ + H${^+}$)~-~N(H$^{0}$) $\sim$ N(S II)$\times${H/S}$_{\odot}$ - N(H$^{0}$). We find H II/(H I + H II)$\sim$ 7\%.

Table \ref{depletions} compares abundances toward GD 246 with solar photosphere abundances. We find there is an indication that N, O, Si, P, Ar, and Fe are depleted from typical solar abundances, but the uncertainties preclude a definitive statement. \citet{2002ApJS..140....3M} and \citet{2002SSRLinsky} have noted the evidence for 20 -- 25\% depletion of O I by dust in the near ISM. For all species listed above part of this depletion might be explained by not taking into account higher ionization states when calculating their gas-phase abundances. N and Si are the only species for which we could measure some of this contribution; however we have only a lower limit on N(N II) and Si III is blended with stellar Si III.

\subsection{Summary}
We have obtained column densities of H I, D I, C II*, N I, O I, Si II, P II, S II, Ar I, and Fe II for the line of sight to the white dwarf GD 246. This is the fourth line of sight with an accurate D/H measurement with \fuse~that probes the region outside the LIC but inside the Local Bubble.
Based on a constant D/H ratio of 1.5$\EE{-5}$ inside the LIC we find a similar ratio outside the LIC. In addition, our sightline-averaged D I/H I measurement for GD 246 is in agreement with the weighted mean of D I/H I reported by \citet{2002ApJS..140....3M} of D I/H I~=~(1.52~$\pm$~0.08)$\EE{-5}$, for the local ISM. For the WD 2331$-$475, HZ 21, and Lan 23 sightlines the column densities of D I, N I, O I, and other species are also measured and ratios are computed. We have used results from all four sightlines to compute revised \fuse~ratios for the local ISM.

\acknowledgments

This work is based on data obtained for the Guaranteed Time Team by the NASA-CNES-CSA \fuse~mission operated by The Johns Hopkins University. Financial support to U. S. participants has been provided in part by NASA contract NAS5-32985 to Johns Hopkins University.  Support for French participation in this study has been provided by CNES.
Based on observations made with the NASA/ESA Hubble Space Telescope, obtained from the Data Archive at the Space Telescope Science Institute, which is operated by the Association of Universities for Research in Astronomy, Inc., under NASA contract NAS 5-26555. These observations are associated with proposal 7296.
The the profile fitting procedure, Owens.f, used in this work was developed by M. Lemoine and the French \fuse~Team.

%\appendix

%\section{Appendicial material}

\bibliography{ms}
\bibliographystyle{apj}
%\nocite{*}

\clearpage

\clearpage

\clearpage

\clearpage

\clearpage

\clearpage

\clearpage

\clearpage

\clearpage

\clearpage

%\begin{figure}
%\plotone{HI_models_paper.ps}
%\caption{Top panel: H I \lya~profile overplotted on stellar models obtained with different values of T$_{eff}$ and log $g$. Bottom panel: blow up of the core of \lya~to better see the differences in the stellar models (in the two panels the data were binned by 4 for display purposes only). \label{lymanmodels}}
%\end{figure}
%\clearpage

\clearpage

\clearpage

\clearpage

%$$$$$$$$$$$$$$$$$$$$$$$$$$$$$$$$$$$$$$$$$$$$$$$$$$$$$$$$$$$$$$$$$$$$$$$$$$$$$ $
\begin{deluxetable}{lcccc}
%\tablecolumns{4}
\tablewidth{0pc}
\tablecaption{Stellar properties \label{stellar_prop}}
\tablehead{ 
\colhead{Quantity} & \colhead{GD 246} & \colhead{WD 2331$-$475}& \colhead{HZ 21} &\colhead{Lan 23}} 
\startdata
WD number		& 2309+105		& 2331$-$475		& 1211+332		& 2247+583		\\	
Spectral Type  		& DA 			& DA			& DO 			& DA			\\
$l$ (deg)			& 87.25			& 334.85		& 175.04	 	& 107.64		\\
$b$ (deg)			& $-$45.11		& $-$64.81		& +80.03		& $-$0.64 		\\
$d$$^{a}$ (pc) 		& $79~\pm~24$ 		& $82~\pm~25$		& $115~\pm~35$  	& $122~\pm~37$		\\
V 			& 13.09 		& 13.1			&14.22 			& 14.26			\\
$T_{\rm{eff}}$ (K)$^{b}$ 	& $53,088~\pm~968$ 	& $51,800~\pm~800$	& $53,000~\pm~2650$	& $59,360~\pm~800$	\\
log~$g$ (cm s$^{-2}$)$^{b}$ & $7.85~\pm~0.07$ 	& $7.79~\pm~0.07$	& $7.8~\pm~0.3$  	& $7.84~\pm~0.05$	\\
Reference		& 1			& 2			& 3			& 2			\\	
\enddata
\tablenotetext{a}{All distances are photometric, 30\%~error adopted.}
\tablenotetext{b}{The uncertainties quoted are 1~$\sigma$.}
\tablerefs{(1) \citet{1999ApJ...517..399N}; (2) \citet{1997ApJ...480..714V}; (3) \citet{1996A&A...314..217D}}
%\tablenotetext{b}{From Vennes et al. 1997}
%\tablenotetext{c}{From Dreizler \& Werner 1996}
\end{deluxetable}
\clearpage
%$$$$$$$$$$$$$$$$$$$$$$$$$$$$$$$$$$$$$$$$$$$$$$$$$$$$$$$$$$$$$$$$$$$$$$$$$$$$$$ 
\begin{deluxetable}{lccccccc}
%\tablecolumns{8}
\tablewidth{0pc}
\tablecaption{Log of \fuse~observations \label{fuse_obs}}
\tablehead{ 
\colhead{Star} & \colhead{Program ID} & \colhead{Date} & \colhead{Aperture$^a$} & \colhead{Exposures} & \colhead{Exp. Time} & \colhead{Mode$^b$} &\colhead{S/N$^c$}\\
\colhead{} & \colhead{} & \colhead{} & \colhead{} & \colhead{} & \colhead{(ksec)} & \colhead{} &\colhead{}}
\startdata
GD 246	& M1010602  & 1999 Dec 09 & MDRS &  2 	& 1.2 	& HIST & 4.6	\\
	& M1010604  & 1999 Dec 10 & LWRS &  7 	& 3.4	& HIST & 6.4	\\
	& P1044101  & 2000 Jul 19 & LWRS & 28 	& 14.8	& HIST & 8.6	\\
	& M1010601  & 2000 Nov 12 & LWRS & 3	& 1.6	& HIST & 5.2	\\
      	& P2042401  & 2001 Jul 14 & MDRS & 51 	& 24.6 	& HIST & 8.3	\\
\cline{1-8}\\
WD 2331$-$475 & P1044202 & 1999 Nov 08 & LWRS & 29 & 20.2 & HIST &	8.2\\
	   & P1044201 & 2000 Jun 23 & LWRS & 20 & 19.4 & HIST &	8.4 \\
           & P1044203 & 2002 Jul 27 & MDRS & 72 & 32.1 & HIST &	8.3 \\
\cline{1-8}\\
HZ 21 & M1080201 & 2000 May 10 & LWRS & 5 & 4.9 & TTAG 	& 4.4 \\
      & P2040801 & 2001 Jan 27 & LWRS & 4 & 12.4 & TTAG & 5.2 \\
      & P2040802 & 2001 Feb 6 & MDRS & 25 & 16.7 & TTAG & 5.5\\
\cline{1-8}\\
Lan 23 &P2510101 & 2001 Jul 19 & LWRS & 8 & 20.4 & TTAG	& 4.7	\\
\enddata
\tablenotetext{a}{See text for definition of aperture.}
\tablenotetext{b}{See text for definition of observing mode.}
\tablenotetext{c}{S/N ratios, per detector pixel, measured in the LiF 1A segment, between 1000--1005~\AA~for all stars except Lan 23, where we use the range 995 -- 1000~\AA.}
\end{deluxetable}
\clearpage
%$$$$$$$$$$$$$$$$$$$$$$$$$$$$$$$$$$$$$$$$$$$$$$$$$$$$$$$$$$$$$$$$$$$$$$$$$$$$$$

\begin{deluxetable}{lccccc}
%\tablecolumns{6}
\tablewidth{0pc}
\tablecaption{STIS echelle observations of GD 246 \label{STIS_obs}}
\tablehead{ 
\colhead{Program ID} & \colhead{Date} & \colhead{Grating} & \colhead{Aperture} & \colhead{Exp. Time} & \colhead{Wavelengths Covered} \\
\colhead{} & \colhead{} & \colhead{} & \colhead{} & \colhead{(s)} &\colhead{(\AA)}}
\startdata
O4G102010  & 1998 Nov 20 & E140M & $0\farcs02\times0\farcs06$ & 1699.0 & 1140--1735 \\
O4G102020  & 1998 Nov 20 & E140H & $0\farcs02\times0\farcs09$ & 2420.0 & 1170--1372 \\
\enddata
\end{deluxetable}
\clearpage

%$$$$$$$$$$$$$$$$$$$$$$$$$$$$$$$$$$$$$$$$$$$$$$$$$$$$$$$$$$$$$$$$$$$$$$$$$$$$$ 

\begin{deluxetable}{lcccccc}
%\tablecolumns{3}
\tablewidth{0pc}
\tablecaption{Atomic data and analysis methods for the lines used in the analyses$^{a}$ \label{atomicdata}}
\tablehead{ 
\colhead{Species} & \colhead{Wavelength (\AA)} & \colhead{Log $f\lambda$} & \colhead{GD 246} &\colhead{WD 2331$-$475}& \colhead{HZ 21} & \colhead{Lan 23}}
\startdata
H I	& 1215.670	& 2.70	& P     &\ldots		& \ldots & \ldots    \\
D I	& 920.713	& 0.17	& \ldots &\ldots	& \ldots & P	\\
\ldots	& 922.899	& 0.31	& \ldots &\ldots	& \ldots & P	 \\
\ldots	& 925.974	& 0.47	& P	 &\ldots	& \ldots & P	 \\
\ldots	& 930.495	& 0.65	& P	 &\ldots	& P	  & P	 \\
\ldots	& 937.548	& 0.86	& P	 & P		& P	  & P	 \\
\ldots	& 949.485	& 1.12	& P	 & P		& P	  & \ldots \\
\ldots	& 972.272	& 1.45	& P	 & P		& \ldots & \ldots \\
\ldots	& 1025.443	& 1.91	& \ldots & P		& \ldots & \ldots \\
C II*	& 1037.018	& 2.11	& P	& P, A		& A	 & A	 \\
\ldots	& 1335.663	& 1.23	& P	&\ldots		& \ldots & \ldots \\
\ldots	& 1335.708	& 2.23	& P	&\ldots		& \ldots & \ldots \\
N I     & 952.303	& 0.25	& \ldots &\ldots	& \ldots & P	\\
\ldots  & 952.415	& 0.21	& \ldots &\ldots	& \ldots & P	 \\
\ldots  & 952.523	& $-$0.24	& \ldots &\ldots	& \ldots & P	\\
\ldots  & 953.415	& 1.10	& \ldots & P, C		& P, C   & \ldots \\
\ldots	& 953.655	& 1.38	& \ldots & P, C		& P, C   & \ldots  \\
\ldots	& 953.9699	& 1.52  & \ldots & P, C		& \ldots & \ldots \\
\ldots  & 954.1042	& 0.81	& \ldots & P		& \ldots & \ldots \\
\ldots	& 963.990	& 1.54	& \ldots & P		& P, C   & \ldots \\
\ldots	& 964.626	& 0.96	& P, C	 & P, C		& P, C   & \ldots \\
\ldots	& 965.041	& 0.59	& P, C	 & P	   	& P, C   & \ldots \\
\ldots  & 1134.1653	& 1.69	& \ldots & P, C		& \ldots & \ldots \\
\ldots  & 1134.4149	& 1.53  & \ldots & P, C		& \ldots & \ldots \\
\ldots  & 1134.9803	& 1.24  & \ldots & P, C		& \ldots & \ldots \\
\ldots	& 1199.550	& 2.19	& C	 &\ldots	& \ldots & \ldots \\
\ldots	& 1200.223	& 2.01	& C	 &\ldots	& \ldots & \ldots \\
\ldots	& 1200.710	& 1.71	& C	 &\ldots	& \ldots & \ldots \\
N II	& 1083.994	& 2.10	& A	 & A		& A	  & A     \\
O I	& 919.658	& $-$0.06	& P, C 	& P, C		& P, C   & \ldots  \\
\ldots	& 919.917	& $-$0.79 & P, C	& P		& P, C	  & P, A \\
\ldots	& 921.875	& 0.04	& P, C	& P, C		& \ldots & \ldots\\
\ldots	& 922.200	& $-$0.65	& P	& P		& \ldots & \ldots\\
\ldots	& 924.950	& 0.15	& P, C	& P, C		& P, C   & \ldots  \\
\ldots	& 925.446	& $-$0.49	& P, C	& P		& P, C   & \ldots \\
\ldots  & 929.5168      & 0.32  &\ldots	& P, C		& \ldots & \ldots \\
\ldots	& 930.257	& $-$0.30	& P, C	& P		& P, C   & \ldots \\
\ldots	& 936.6295	& 0.53	&\ldots	& P, C		&\ldots  &\ldots \\
\ldots	& 948.6855	& 0.77	& C	& C		& C	 & C	\\
\ldots	& 950.885	& 0.18	& P, C	&\ldots		& \ldots & \ldots \\
\ldots	& 971.738	&1.13	& C	& C		& C	 & C	 \\
\ldots  & 972.142	& $-$0.47 & P	& P		& \ldots & \ldots \\
\ldots	& 974.070	& $-$1.82	&\ldots &\ldots		& \ldots & P \\
\ldots	& 976.4481	& 0.51	&\ldots & P, C		&\ldots	 &\ldots \\	
\ldots  & 1039.2301	&0.98	& C	& C		& C	 & C	\\
Si II	& 1020.699	& 1.22	& P, C	& P, A		& P, A   & A	  \\
\ldots	& 1190.416	& 2.54	& P, C	&\ldots		& \ldots & \ldots \\
\ldots	& 1193.230	& 2.84	& P, C	&\ldots		& \ldots & \ldots \\
\ldots	& 1304.370	& 2.08	& P, C	&\ldots		& \ldots & \ldots \\
P II	& 963.800	& 3.15	& P	& P, A		& P, A   & A	 \\
S II	& 1250.584	& 0.83	& P, C	&\ldots		& \ldots & \ldots \\
\ldots	& 1253.811	& 1.14	& P, C	&\ldots		& \ldots & \ldots \\
\ldots	& 1259.519	& 1.32	& P, C	&\ldots		& \ldots & \ldots \\
Ar I	& 1048.220	& 2.44	& P	& P, A		& P, A   & A	 \\
Fe II	& 1063.176	& 1.76	& P	&\ldots		& P, A	  & P, C	\\
\ldots	& 1121.975	& 1.36	&\ldots &\ldots		& \ldots & P, C \\
\ldots	& 1125.448	& 1.26	&\ldots &\ldots		& \ldots & P, C, A \\
\ldots	& 1144.938	& 2.08	& P	& P, A		& \ldots & P, C\\
\enddata
\tablenotetext{a}{P, C, and A, denote lines that are analyzed with profile fitting, curve of growth, and apparent optical depth, respectively.}
\end{deluxetable}
\clearpage

\begin{deluxetable}{lccc}

%$$$$$$$$$$$$$$$$$$$$$$$$$$$$$$$$$$$$$$$$$$$$$$$$$$$$$$$$$$$$$$$$$$$$$$$$$$$$$ 

\begin{deluxetable}{lcccc}
%\tablecolumns{4}
\tablewidth{0pc}
\tablecaption{Adopted column densities$^{a}$ \label{cols}}
\tablehead{ 
\colhead{Species} & \colhead{GD 246 (cm$^{-2}$)} & \colhead{WD 2331$-$475 (cm$^{-2}$)}&\colhead{HZ 21 (cm$^{-2}$)} & \colhead{Lan 23$^b$ (cm$^{-2}$)}}
\startdata
log N(H I) 	& $19.11~\pm~0.05$ 	 & \ldots	   	& \ldots		& $19.89~\pm~^{0.49}_{0.08}$\\
log N(D I) 	& $14.29~\pm~0.09$ 	 & $14.19~\pm~0.12$	& $14.40~\pm~0.15$	& $15.23~\pm~0.13$  	\\
log N(C II*) 	& $13.05~\pm~0.04$ 	 & $13.24~\pm~0.17$	& $12.97~\pm~^{0.20}_{0.27}$ 	&  $\geq~13.65$			\\
log N(N I)  	& $14.75~\pm~0.06$ 	 & $14.53~\pm~0.10$	& $14.77~\pm~0.08$	& $15.73~\pm~^{0.13}_{0.15}$ \\
log N(N II)	& $\geq~13.81$		 & $\geq~14.27$			&$\geq~14.03$			& $\geq~13.95$	\\
log N(O I)  	& $15.67~\pm~0.07$	 & $15.48~\pm~0.11$	& $15.74~\pm~0.10$	& $16.72~\pm~^{0.29}_{0.38}$\\
log N(Si II) 	& $14.07~\pm~^{0.06}_{0.04}$ 	 & $13.99~\pm~0.20$	& $14.36~\pm~0.08$    & $\geq~14.52$   \\
log N(P II) 	& $12.29~\pm~0.10$ 	 & $12.18~\pm~0.20$	& $12.57~\pm~^{0.12}_{0.15}$	& $\geq~12.72$      \\
log N(S II) 	& $14.34~\pm~0.02$ 	 &  \ldots			&  \ldots			& \ldots\\
log N(Ar I) 	& $13.14~\pm~^{0.13}_{0.10}$   	 & $12.77~\pm~0.20$	& $13.16~\pm~0.05$	& $\geq~13.50$      \\
log N(Fe II)	& $13.30~\pm~0.10$ 	 & $13.32~\pm~0.10$	& $13.54~\pm~0.08$	&$14.03~\pm~0.13$ \\
\enddata
\tablenotetext{a}{Adopted values are the combination of PF1 and PF2 (see text for more on this). All uncertainties are 2$\sigma$.}
\tablenotetext{b}{N(H I) from \citet{1999A&A...346..969W}. Range quoted includes uncertainty in the photospheric composition of this star (see \S 5).}
\end{deluxetable}
\clearpage

%$$$$$$$$$$$$$$$$$$$$$$$$$$$$$$$$$$$$$$$$$$$$$$$$$$$$$$$$$$$$$$$$$$$$$$$$$$$$$$

%\tablecolumns{3}
\tablewidth{0pc}
\tablecaption{H$_{2}$ column densities in Lan 23$^{a}$ \label{h2cols}}
\tablehead{ 
\colhead{logN(H$_{2}$)} & \colhead{Profile Fitting$^{b}$} & \colhead{Curve of Growth$^{c}$}}
\startdata
$J~=~$0 & 14.35~$\pm~^{0.05}_{0.09}$ & 14.12~$\pm~^{0.08}_{0.09}$ \\ %& 14.24$^{0.05}_{0.05}$  \\
$J~=~$1 & 14.97~$\pm~^{0.07}_{0.09}$ & 14.89~$\pm~0.11$ \\%& 14.94$^{0.06}_{0.07}$  \\
$J~=~$2 & 14.32~$\pm~^{0.09}_{0.13}$ & 14.20~$\pm~^{0.08}_{0.05}$ \\%& 14.23$^{0.06}_{0.06}$  \\
$J~=~$3 & 13.95~$\pm~0.07$ & 13.93~$\pm~^{0.11}_{0.10}$ \\%& 13.94$^{0.06}_{0.07}$  \\
\enddata
\tablenotetext{a}{All uncertainties are 2$\sigma$.}
\tablenotetext{b}{The $b$ value of the fit was constrained to 4.1~km~s$^{-1}$. The errors quoted do not include a contribution from errors in the $b$ value.}
\tablenotetext{c}{All lines are fit simultaneously yielding $b$~=~4.1$^{+0.8}_{-0.6}$~km~s$^{-1}$ (2$\sigma$).}
\end{deluxetable}
\clearpage

%$$$$$$$$$$$$$$$$$$$$$$$$$$$$$$$$$$$$$$$$$$$$$$$$$$$$$$$$$$$$$$$$$$$$$$$$$$$$$$
\begin{deluxetable}{lcccccc}
%\tablecolumns{3}
\tablewidth{0pc}
\tablecaption{Ratios of column densities$^a$ \label{ratios}}
\tablehead{ 
\colhead{Quantity} & \colhead{GD 246} &\colhead{WD 2331$-$475} & \colhead{HZ 21} & \colhead{Lan 23}}
\startdata
D I/H I~($\times10^{-5}$) & $1.51~\pm~^{0.39}_{0.33}$ &\ldots &\ldots & $2.19~\pm~^{4.64}_{0.68}$  \\
O I/H I~($\times10^{-4}$) & $3.63~\pm~^{0.77}_{0.67}$ &\ldots &\ldots & $6.76~\pm~^{15.5}_{4.10}$ \\
D I/O I~($\times10^{-2}$) & $4.17~\pm~^{1.21}_{1.00}$ &$5.13~\pm~^{2.20}_{1.69}$& $4.57~\pm~^{2.22}_{1.63}$ & $3.24~\pm~^{3.27}_{2.06}$  \\
N I/H I~($\times10^{-5}$) & $4.37~\pm~^{0.84}_{0.74}$ 	&  \ldots &  \ldots	& $6.92~\pm~^{14.7}_{2.33}$ \\
D I/N I~($\times10^{-1}$) & $3.47~\pm~^{0.95}_{0.79}$ &$4.57~\pm~^{1.88}_{1.45}$	& $4.27~\pm~^{1.96}_{1.44}$ & $3.16~\pm~^{1.56}_{1.23}$  \\
O I/N I	& $8.32~\pm~^{1.91}_{1.64}$  & $8.91~\pm~^{3.45}_{2.71}$ & $9.33~\pm~^{3.07}_{2.48}$& $9.77~\pm~^{9.89}_{6.37}$ \\
%Si//H & $1.51~\pm~0.37~\times~10^{-5}$ & &\\
%P/H & $1.51~\pm~0.37~\times~10^{-5}$ & &\\
%S/H & $1.51~\pm~0.37~\times~10^{-5}$ & &\\
%Ar/H & $1.51~\pm~0.37~\times~10^{-5}$ & &\\
%Fe/H & $1.51~\pm~0.37~\times~10^{-5}$ & &\\
%D/O & $4.17~\pm~1.20~\times~10^{-2}$ & &\\
%D/N & $3.47~\pm~0.95~\times~10^{-1}$ & &\\
\enddata
\tablenotetext{a}{All uncertainties are 2$\sigma$.}
\end{deluxetable}
\clearpage
%$$$$$$$$$$$$$$$$$$$$$$$$$$$$$$$$$$$$$$$$$$$$$$$$$$$$$$$$$$$$$$$$$$$$$$$$$$$$$$
\begin{deluxetable}{lcccc}
\tablewidth{0pc}
\tablecaption{Revised \fuse~ratios and comparison with other measurements$^{a}$ \label{comparison}}
\tablehead{ 
\colhead{Quantity} & \colhead{\citet{2002ApJS..140....3M}}& \colhead{$\chi^{2}_{\nu}$ for}&\colhead{Degrees of} & \colhead{Meyer et al.}\\
\colhead{} & \colhead{+ this work$^b$} & \colhead{mean} & \colhead{freedom $\nu^c$} & \colhead{(1997, 1998)$^d$}}
\startdata
D I/H I & $1.52~\pm~0.07$~ (0\%)	&1.0	&5 (4)	&\ldots			\\
($\times10^{-5}$) & &  &  &\\
O I/H I & $3.17~\pm~0.19$~(+5\%)	&3.5	&5 (4)	& 3.43$~\pm~$0.15	\\
($\times10^{-4}$) &  & &  &\\
D I/O I & $4.06~\pm~0.17$~(+2\%)	&1.9	&9 (6)	&\ldots			\\
($\times10^{-2}$) & &  &  &\\
N I/H I& $4.28~\pm~0.25$~(+1\%)		&1.4	&4 (3)	&7.5$~\pm~$0.4		\\
($\times10^{-5}$) & &  & &\\
D I/N I & $3.41~\pm~0.15$~(+1\%)	&2.2	&9 (5)	&\ldots			\\
($\times10^{-1}$) & &  & &\\
O I/N I	 & $8.23~\pm~0.38$~(+2\%)	&2.9	&8 (5)	& 4.6$~\pm~$0.3		\\
\enddata
\tablenotetext{a}{{\it All uncertainties in this table are 1$\sigma$~in the mean.}}
\tablenotetext{b}{Mean values and uncertainties in the mean. The values in () correspond to the percent increase from the Moos et al. (2002) values. Our values are combined with those from \citet{2002ApJS..140....3M} by taking a weighted mean. We use the largest of the lower and upper error bars of each individual ratio to compute the weighted mean.}
\tablenotetext{c}{$\nu$~=~number of sightlines~$-$~1. Values in () represent the number of degrees of freedom in \citet{2002ApJS..140....3M}}
\tablenotetext{d}{The values in the table correspond to O/H and N/H, the uncertainties quoted are errors in the mean. O/N calculated here from O/H and N/H.}
\end{deluxetable}
\clearpage
%$$$$$$$$$$$$$$$$$$$$$$$$$$$$$$$$$$$$$$$$$$$$$$$$$$$$$$$$$$$$$$$$$$$$$$$$$$$$$$

\begin{deluxetable}{lccc}
%\tablecolumns{3}
\tablewidth{0pc}
\tablecaption{Comparison of abundances toward GD 246 with solar abundances \label{depletions}}
\tablehead{ 
\colhead{X} & \colhead{log(X/H)$_{\odot}$~+12}  & \colhead{log(X/H)$^{a}_{GD~246}$} & \colhead{Reference}}
\startdata
N	& 7.93 & 7.64~$\pm~^{1.30}_{1.30}$	& 1	\\
O	& 8.69 & 8.56~$\pm~^{1.45}_{1.45}$	& 2	\\
Si 	& 7.54 & 6.96~$\pm~^{1.28}_{1.11}$	& 1	\\
P	& 5.51$^{b}$ & 5.18~$\pm~^{1.56}_{1.56}$	& 3	\\
S	& 7.27$^{b}$ & 7.23~$\pm~^{1.00}_{1.00}$	& 3	\\
Ar	& 6.40 & 6.03~$\pm~^{1.96}_{1.62}$	& 3	\\
Fe	& 7.45 & 6.19~$\pm~^{1.64}_{1.64}$	& 1	\\
\enddata
\tablenotetext{a}{All uncertainties are 2$\sigma$.}
\tablenotetext{b}{The meteoritic and photospheric values were averaged.}
\tablerefs{(1) \citet{2001sgc..conf...23H}; (2) \citet{2001ApJ...556L..63A}; (3) \citet{1998SSRv...85..161G}.}
\end{deluxetable}
\clearpage

\end{document}